\documentclass[sigconf]{acmart}

\setcopyright{none}

\renewcommand\footnotetextcopyrightpermission[1]{} % removes footnote with conference information in first column
\makeatletter
\renewcommand\@formatdoi[1]{\ignorespaces}
\makeatother

\newcommand\IsoApprCount{1600}

\newcommand\IsoApprSizeGB{1843 GB}

\newcommand\DebApprCount{79000}

\newcommand\DebApprSizeGB{36 GB}

\newcommand\ELFApprCount{105000}

\newcommand\ELFApprSizeGB{29 GB}

\newcommand\ELFExaCount{105013}

\newcommand\ELFExaSizeGB{29.74 GB}

\newcommand\CodesecApprCount{105000}

\newcommand\CodesecApprSizeGB{17 GB}

\newcommand\SamplesAllExaCountClemens{16785}
\newcommand\SamplesAllExaCountDeNicolao{15290}
\newcommand\SamplesAllExaCount{67285}

\newcommand\ArchsExaCountClemens{20}
\newcommand\ArchsExaCountDeNicolao{21}
\newcommand\ArchsExaCount{23}

\newcommand\ArchSetSizeAvgApprCount{3000}
\newcommand{\rpm}{\raisebox{.2ex}{$\scriptstyle\pm$}}

\newcommand{\specialcell}[2][c]{%
	\begin{tabular}[#1]{@{}c@{}}#2\end{tabular}}

\usepackage{color}
\usepackage{framed}

\newcounter{t0d0_counter}
\newcommand{\nofixme}[1]{
}
\newcommand{\fixme}[1]{
  \stepcounter{t0d0_counter}
  \definecolor{shadecolor}{rgb}{1,1,0} % this is yellow
  \begin{shaded}
  T0D0 \arabic{t0d0_counter}: #1
  \end{shaded}
}

\AtBeginDocument{%
  \providecommand\BibTeX{{%
    \normalfont B\kern-0.5em{\scshape i\kern-0.25em b}\kern-0.8em\TeX}}}

\usepackage{graphicx}
\usepackage{comment}
\usepackage{capt-of}
\usepackage{booktabs}
\usepackage{hyperref}
\usepackage{enumitem}
\usepackage{listings}

\acmConference{arXiv.org}{pre-print}{15.8.2019}
\begin{document}

\title{Towards usable automated detection of CPU architecture and endianness for arbitrary binary files and object code sequences}


%\author{<Blinded For Review>}

\author{
Sami Kairaj\"arvi
}
\authornote{This paper is based on author's MSc thesis~\cite{Kairajarvi:Thesis:2019}.}
\affiliation{%
  \institution{University of Jyv\"askyla, Jyv\"askyla, Finland}
}
\email{samakair@jyu.fi}
%%\orcid{1234-5678-9012}
%%\authornotemark[1]
%%\email{webmaster@marysville-ohio.com}
  %%\streetaddress{P.O. Box 1212}
  %%\city{Dublin}
  %%\state{Ohio}
  %%\postcode{43017-6221}

\author{Andrei Costin}
\affiliation{%
  \institution{University of Jyv\"askyla, Jyv\"askyla, Finland}
}
\email{ancostin@jyu.fi}
\orcid{0000-0002-2704-9715}

\author{Timo H\"am\"al\"ainen}
\affiliation{%
  \institution{University of Jyv\"askyla, Jyv\"askyla, Finland}
}
\email{timoh@jyu.fi}
\orcid{0000-0002-4168-9102}

\renewcommand{\shortauthors}{S. Kairaj\"arvi, A. Costin, T. H\"am\"al\"ainen}
%\renewcommand{\shortauthors}{<Blinded For Review>}

\begin{abstract}
Static and dynamic binary analysis techniques are actively %and successfully 
used to reverse engineer software's behavior and to detect its vulnerabilities, 
even when only the binary code is available for analysis. 
To avoid analysis errors due to misreading op-codes for a 
wrong CPU architecture, these analysis tools must precisely identify the 
Instruction Set Architecture (ISA) of the object code under analysis. 
%Sometimes this information is available in most common binary file format headers, 
%but many other times this information could be either missing or wrong. 
%
The variety of CPU architectures that modern security and reverse 
engineering tools must support is ever increasing due to massive 
proliferation of IoT devices and the diversity of firmware and malware 
targeting those devices. 
Recent studies concluded that falsely identifying the binary code's 
ISA caused alone about 10\% of failures of IoT firmware analysis. %~\cite{costin2014large,costin2016automated}. 
The state of the art approaches to detect ISA for arbitrary object code look 
promising -- their results demonstrate effectiveness and 
high-performance. %~\cite{clemens2015automatic,de2018elisa}. 
However, they lack the support of publicly available datasets 
and toolsets, which makes the evaluation, comparison, and improvement of 
those techniques, datasets, and machine learning models quite 
challenging (if not impossible). 
This paper bridges multiple gaps in the field of automated and precise 
identification of architecture and endianness of binary files and object code. 
We develop from scratch the toolset and datasets that are lacking 
in this research space. As such, we contribute a comprehensive collection 
of \emph{open data}, \emph{open source}, and \emph{open API} web-services. 
%
%Using those toolsets and datasets 
We also attempt experiment reconstruction and cross-validation of 
effectiveness, efficiency, and results of the state of the art methods. %by Clemens~\cite{clemens2015automatic} and De Nicolao et al.~\cite{de2018elisa}. 
%
%Therefore, an experimental evaluation was performed to test classifier 
%performance in multiple different scenarios. 
%
When training and testing classifiers using solely code-sections 
from executable binary files, all our classifiers performed equally well 
achieving over 98\% accuracy. 
%On samples with very small code sections 
%3-NN and SVM perfomed best achieving 90\% accuracy at 128 bytes. 
%
The results are consistent and comparable with the current state of 
the art, hence supports the general validity of the algorithms, features, 
and approaches suggested in those works. 
Complementing the field, we propose using complete binaries in either testing or 
training\&testing mode -- experiments show that ISA of complete binary files is identified 
%with 90\% accuracy when Random Forest classifier trained only on code-sections, and 
with 99.2\% accuracy using Random Forest classifier. 
%trained on complete binaries. 
\end{abstract}

% TODO




% AIsec'19
% https://www.acm.org/binaries/content/assets/publications/article-templates/acm-update.pdf
\maketitle

%%%%%%%%%%%%%%%%%%%%%%%%%%%%%%%%%%%%%%%%%%%%%%%%%%%%%%%%%%%%%%%%%%%%%%%%%%%%%%%%

\section{Introduction}
\label{sec:introduction}

Reverse engineering and analysis of binary code has a wide spectrum of applications~\cite{sutherland2006empirical,liu2013binary,shoshitaishvili2016sok}, 
ranging from vulnerability research~\cite{wang2009intscope,cha2012unleashing,costin2014large,shoshitaishvili2015firmalice,costin2016automated} to binary 
patching and translation~\cite{sites1993binary}, and from digital 
forensics~\cite{clemens2015automatic} to anti-malware and 
Intrusion Detection Systems (IDS)~\cite{van2009detecting,manni2014network}. 
For such applications, various static and dynamic analysis techniques and tools 
are constantly being researched, developed, and 
improved~\cite{song2008bitblaze,chipounov2011s2e,brumley2011bap,liu2013binary,wang2017angr,radare2,eagle2011ida}. 

Regardless of their end goal, one of the important steps in these techniques 
is to correctly identify the Instruction Set Architecture (ISA) of the op-codes 
within the binary code. Some techniques can perform the analysis using  
architecture-independent or cross-architecture methods~\cite{pewny2015cross,eschweiler2016discovre,feng2016scalable}. 
However, many of those techniques still require the exact knowledge of 
the binary code's ISA. For example, recent studies concluded that falsely 
identifying the binary code's ISA caused about 10\% of failures of 
IoT/embedded firmware analysis~\cite{costin2014large,costin2016automated}. 

Sometimes the CPU architecture is available in the executable format's 
header sections, for example in ELF file format~\cite{nohr1993unix}. 
However, this information is not guaranteed to be universally available
for analysis. There are multiple reasons for this and we will detail a few of them. 
The advances in Internet of Things (IoT) technologies bring 
to the game an extreme variety of hardware and software, in particular 
new CPU architectures, new OSs or OS-like systems~\cite{muench2018you}. 
Many of those devices are resource-constrained and the binary code comes 
without sophisticated headers and OS abstraction layers. 
At the same time, the digital forensics and the IDSs sometimes may have access 
only to a fraction of the code, e.g., from an exploit (shell-code), 
malware trace, or a memory dump. For example, many shell-codes are exactly this 
-- a quite short, formatless and headerless sequence of CPU op-codes for a target 
system (i.e., a combination of hardware, operating-system, and abstraction 
layers) that performs a presumably malicious action on behalf of the 
attacker~\cite{foster2005sockets,polychronakis2010comprehensive}. 
In such cases, though possible in theory, it is quite unlikely that the full 
code including the headers specifying CPU ISA will be available for analysis. 
Finally, even in the traditional computing world of (e.g., \texttt{x86/x86\_64}), 
there are situations where the hosts contain object code for CPU architectures 
other than the one of the host itself. Examples include firmware for network 
cards~\cite{delugre2010closer,duflot2011if,blanco2012one}, various management 
co-processors~\cite{miller2011battery}, and device 
drivers~\cite{chipounov2010reverse,kadav2012understanding} for 
USB~\cite{nohl2014badusb,tian2015defending} 
and other embedded devices that contain own specialized processors and provide 
specific services (e.g., encryption, data codecs) that are implemented as 
peripheral firmware~\cite{li2011viper,davidson2013fie}. 
Even worse, more often than not the object code for such peripheral firmware 
is stored using less traditional or non-standard file formats and headers 
(if any), or embedded inside the device drivers themselves resulting 
in mixed architectures code streams. 

Currently, several state of the art works try to address the challenge of 
accurately identifying the CPU ISA for arbitrary object code 
sequences~\cite{clemens2015automatic,de2018elisa}. Their approaches look 
promising as the results demonstrate effectiveness and high-performance. 
However, they lack the support of publicly available datasets and toolsets, 
which makes the evaluation, comparison, and improvement of those techniques, 
datasets, and machine learning models quite challenging (if not impossible). 

With this paper, we bridge multiple gaps in the field of automated and precise 
identification of architecture and endianness of binary files and object code. 
We develop from scratch the toolset and datasets that are lacking 
in this research space. To this end, we release a comprehensive collection 
of \emph{open data}, \emph{open source}, and \emph{open API} web-services. 
We attempt experiment reconstruction and cross-validation of effectiveness, 
efficiency, and results of the state of the art 
methods~\cite{clemens2015automatic,de2018elisa}, as well as propose 
and experimentally validate new approaches to the main classification challenge 
where we obtain consistently comparable, and in some scenarios better, results. 
The results we obtain in our extensive set of experiments are consistent and 
comparable with prior art, hence supports the general validity and soundness 
of both existing and newly proposed algorithms, features, and approaches.

\subsection{Contributions}
\label{sec:contribution}
%In this paper, we focus on automated and practical detection of 
%architecture and endianness for arbitrary binary code sequences. 
%Our contribution can be summarized as follows:
In this paper, we present the following contributions:

\begin{enumerate}[label=\alph*)]

\item First and foremost contribution is that 
we implement and release as \emph{open source} the code and toolset 
necessary to reconstruct and re-run the experiments 
from this paper as well as from the state of the art works of 
Clemens~\cite{clemens2015automatic} and De Nicolao et al.~\cite{de2018elisa}.
%~\footnote{
%The release includes: 
%code for collecting and generating datasets and testsets; 
%code for machine learning classification; code for exposing the architecture 
%as Open APIs and its integration in most popular reverse engineering frameworks 
%(e.g., Radare~\cite{radare2})}.
To the best of our knowledge, this is the first such toolset to be publicly 
released. %~\footnote{For the purpose of this submission, web-service and APIs are accessible at: \url{http://34.73.204.185:5000/}. }. 

\item Second and equally important contribution is that 
we release as \emph{open data} the machine learning models and data 
necessary to both validate our results and expand further the datasets and the research field. 
To the best of our knowledge, this is both the first and the largest dataset of 
this type to be publicly released. 

\item Third valuable contribution is that 
we propose, evaluate and validate the use of ``complete binaries'' when training 
the classifiers (as opposed to current ``code-sections''-only 
approaches ~\cite{clemens2015automatic,de2018elisa}). 
We achieve a 99.2\% classification accuracy when testing with Random Forest 
classifiers implemented in Azure Machine Learning platform. 

\item Last but not least, we perform and present the first independent study that 
attempts experiment reconstruction and cross-validation of results, effectiveness 
and efficiency of state of the art methods 
(Clemens~\cite{clemens2015automatic}, De Nicolao et al.~\cite{de2018elisa}). 

\item We release both the dataset and the toolset as open source and open data, 
and are accessible at: 
%\url{https://<Blinded For Review>}
\url{https://github.com/kairis/}

\end{enumerate}

\subsection{Organization}
\label{sec:organization}

% Example: https://link.springer.com/content/pdf/10.1007%2F978-3-319-98989-1.pdf
The rest of this paper is organized as follows.
We review the related work in Section~\ref{sec:relatedwork}. 
Then we detail our methodology, experimental setups and datasets in 
Section~\ref{sec:methodology}. 
We provide a detailed analysis of results in Section~\ref{sec:results}. 
%and discuss specific topics and future work in Section~\ref{sec:discussion}.
%Finally, we conclude with Section~\ref{sec:conclusion}.
Finally, we discuss future work and conclude with Section~\ref{sec:conclusion}.

%%%%%%%%%%%%%%%%%%%%%%%%%%%%%%%%%%%%%%%%%%%%%%%%%%%%%%%%%%%%%%%%%%%%%%%%%%%%%%%%

\section{Related work}
\label{sec:relatedwork}

McDaniel and Heydari~\cite{mcdaniel2003content} introduced the idea to use file contents to identify the file type. Previous methods utilized metadata such as fixed file extension, fixed "magic numbers" and proprietary descriptive file wrappers. The authors showed that many file types have characteristic patterns that can be used to differentiate them from other file formats. They used byte frequency analysis, byte frequency cross-correlation analysis and file header/trailer analysis, and reported accuracies ranging from 23\% to 96\% depending on the algorithm used. A lot of research has utilized the difference in byte frequencies when classifying file types~\cite{fitzgerald2012using,li2011novel,li2005fileprints,xie2013byte,beebe2013sceadan,sportiello2012context,penrose2013approaches}. 
Clemens~\cite{clemens2015automatic} applied the techniques introduced by McDaniel and Heydari~\cite{mcdaniel2003content} and proposed methods for classifying object code with its target architecture and endianness. The author used byte-value histograms combined with signatures to train different classifiers using supervised machine learning. This approach produced promising results and showed that machine learning can be an effective tool when classifying architecture of object code. Classification accuracy varied depending on the used classifier from 92.2\% to 98.4\%. The authors proposed future research ideas for a larger data set, which would include more architectures and code samples compiled with different compilers. 
De Nicolao et al.~\cite{de2018elisa} extend the work of Clemens~\cite{clemens2015automatic} as part of their research by adding more signatures to use as an input for the classifier. They obtain a global accuracy of 99.8\% with logistic regression, which is higher than previously demonstrated by Clemens~\cite{clemens2015automatic}. For further research, they propose to incorporate instruction-level features such as trying to group bytes corresponding to code into valid instructions.
In a slightly different direction, Costin et al.~\cite{costin2016automated} go through every executable inside the firmware and use ELF headers to identify the architecture if the header is present. The authors determine the architecture of the firmware by counting the amount of architecture specific binaries the firmware contains. This information is used to launch an emulator for that specific architecture.
cpu\_rec \cite{cpurec} uses the fact that the probability distribution of one byte depends on the value of the previous one and can detect 72 code architectures~\cite{granboulan2017cpurec}. The author uses Markov chains and Kullback-Leibler divergence for classification. A sliding window is used to handle files with code for multiple architectures in them. The author does not publish any performance measures other than that the analysis of 1 MB takes 60 seconds on 1GB of RAM.
Angr static analysis framework~\cite{shoshitaishvili2016sok} includes the 
Boyscout tool. It leverages static signatures to identify the
CPU architecture executable files. Boyscout tries to match the file to a 
set of signatures containing the byte patterns of function prologues and 
epilogues for known and surveyed architectures. One of the limitations of 
such an approach is that the signatures require maintenance. The performance 
of the classification is highly dependent on keeping those signatures up-to-date, 
complete and highly-qualitative, which can be challenging. Also, this 
technique can be less effective for heavily optimized code which is many 
times the case with the resource-constrained IoT/embedded devices. 
Binwalk \cite{binwalk} uses architecture-specific features (totalling 33 signatures for 9 CPU architectures) along with Capstone disassembler~\cite{quynh2014capstone} (having 9 configurations for 4 CPU architectures) to identify object code's architecture~\footnote{Last checked on 20.4.2019.}. Using only signatures has some limitations as for example they can lead to false positives if the signatures are not unique compared to other architectures in the dataset~\cite{clemens2015automatic}. In addition to that, the disassembly method requires complete support from a disassembler framework, which might not always be available or working perfectly. 
%
%The classifiers proposed in the previous research have good performance regarding the accuracy, but there is a gap of collecting and pre-processing binaries from different architectures for analysis purposes. For example, the \cite{clemens2015automatic} data set is not publicly available and is limited in size. Acquiring such data sets or developing own tools takes a long time, huge human effort, and can be prone to errors. As a part of this thesis, tools to obtain a data set will be provided. This allows researchers to focus on developing, extending and improving the classifiers and not waste valuable time on setting up the infrastructure and acquiring the data set. 
%	
%There is also a need for more comprehensive evaluation of classifier performance on full binaries that includes more classifiers and architectures. \cite{de2018elisa} tested only the performance of logistic regression on a small data set consisting of full binaries. In this regard, the present thesis bridges this gap by conducting a larger experiment, where the classification performance is tested with a large dataset consisting of full binaries for as many architectures as possible. Different classifiers are also used to see if some classifiers perform better than others when classifying full binaries.
%
Costin et al.~\cite{costin2017towards} were the first to apply machine learning in the context of firmware classification. They try to automate finding the brand and the model of the device the firmware is made for and propose different firmware features and machine learning algorithms to get this information. The researchers achieve the best results with random forest classifier with 90\% classification accuracy. With the help of statistical confidence interval, the authors estimate that in a data set of 172 000 firmware images, the classifier could correctly classify the firmware in 93.5\% \rpm 4.3\% of the cases. 
    %The research suggests same kind of techniques could be used to complement firmware vulnerability discovery techniques.

%%%%%%%%%%%%%%%%%%%%%%%%%%%%%%%%%%%%%%%%%%%%%%%%%%%%%%%%%%%%%%%%%%%%%%%%%%%%%%%%

\section{Datasets and experimental setup}
\label{sec:methodology}

\subsection{Datasets}
\label{sec:datasets}
We started with the dataset acquisition challenge. Despite the existence of 
several state of the art works, unfortunately neither their datasets nor 
the toolsets are publicly available.~\footnote{A 
post-processed dataset from Clemens~\cite{clemens2015automatic} 
was generously provided by the author privately on request. The dataset is not publicly available. }
To overcome this limitation we had to develop a complete toolset and 
pipeline that are able to optimally download a dataset that is both large 
and representative enough. We release our data 
acquisition toolset as open source. 
The pipeline used to acquire our dataset is depicted 
in Figure~\ref{fig:workflow}, while sample code snippets are listed 
in Appendix~\ref{apdx:apdx}~\footnote{Due to space limitation, only a part 
of the code is listed in this submission.}. 

	\begin{figure}
		\includegraphics[keepaspectratio,width=\linewidth]{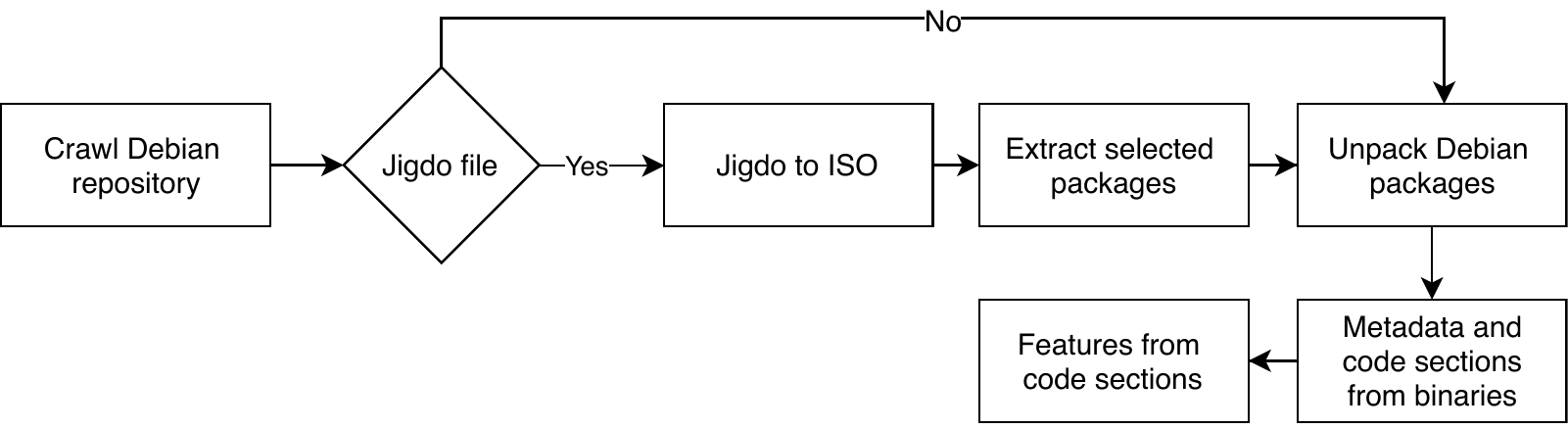}
		\captionof{figure}{The workflow of our toolset to download and prepare the datasets.}
		\label{fig:workflow}
	\end{figure}

We chose the Debian Linux repositories for several reasons. 
First, it is a long established and trusted project, therefore a 
good source of software packages and related data. 
Second, it is bootstrapped from the same tools and sources to 
build the Linux kernel and userspace software packages for a very 
diverse set of CPU and ABI architectures. 

The downloaded and pre-processed dataset can be 
summarized as follows: 
about \IsoApprCount{} ISO/Jigdo files taking up around \IsoApprSizeGB{}; 
approximately \DebApprCount{} DEB package files taking up about \DebApprSizeGB{};
around \ELFApprCount{} ELF files taking up about \ELFApprSizeGB{};
about \CodesecApprCount{} ELF code sections taking up approximately 
\CodesecApprSizeGB{}. A detailed breakdown of our dataset and sample-sets 
is in Table~\ref{table:supported_architectures}.
	\begin{table}
	\captionof{table}{Dataset details in terms of filesystem size and file counts.}
		\label{table:data set_cumulative}
			\begin{tabular}{@{}lrr@{}}
				\toprule
				Type                & Approx. \# of files in dataset & Approx. total size \\ 
				\midrule
				.iso files          & \textasciitilde \IsoApprCount{}      &     \textasciitilde \IsoApprSizeGB{}    \\
				.deb files          & \textasciitilde \DebApprCount{}       &     \textasciitilde \DebApprSizeGB{}        \\
				ELF files           & \ELFApprCount{}      &     \textasciitilde \ELFApprSizeGB{}     \\ 
				ELF code sections   & \CodesecApprCount{}      &     \textasciitilde \CodesecApprSizeGB{}        \\
				\bottomrule
			\end{tabular}
	\end{table}
%
\begin{comment}
%Hurd-i386 architecture was left out, as it uses i386 architecture and is considered a port because of the different kernel it utilizes. AVR and Cuda samples were left out as well as there was no source for sufficiently reliable and representative set of binary files for the architectures. Not every Debian version available for every architecture was downloaded due to resource constraints. Also, only about 67 000 executables were used to train and test the classifier due to resource constraints. The summarized details of the data set are presented in Tables \ref{table:data set_cumulative} and \ref{table:supported_architectures}. 
\end{comment}

% TODO: FIXME:
%\begin{comment}
    % https://docs.google.com/spreadsheets/d/175bJfbwVpULU96yFePuv3ucpu8b8o1Av2lAhr9oN54I/edit?usp=sharing
	\begin{table*}%{\linewidth}
		\captionof{table}{Dataset details and statistics, summarized per architecture.}
		\label{table:supported_architectures}
		\resizebox{\textwidth}{!}{%
			\begin{tabular}{@{}lrcllrr@{}}
				\toprule
				Architecture & \specialcell{Sample-sets size \\ in experiment} & Wordsize & Endianness & File info                                     &  \specialcell{\# of samples \\ in dataset}    & \specialcell{Size of samples \\ in dataset (GB)} \\ 
				\midrule
				alpha        & 3000          & 64       & Little     & ELF 64-bit LSB executable, Alpha (unofficial) &    4042      &    	1.62    \\
				amd64        & 2994          & 64       & Little     & ELF 64-bit LSB executable, x86-64             &    6221      &    	1.19    \\
				arm64        & 2997          & 64       & Little     & ELF 64-bit LSB executable, ARM aarch64        &    4255      &    	0.84    \\
				armel        & 2994          & 32       & Little     & ELF 32-bit LSB executable, ARM                &    4621      &    	0.86    \\
				armhf        & 2994          & 32       & Little     & ELF 32-bit LSB executable, ARM                &    4618      &    	0.70    \\
				hppa         & 3000          & 32       & Big        & ELF 32-bit MSB executable, PA-RISC (LP64)     &    	4909      &    	1.48    \\
				i386         & 2994          & 32       & Little     & ELF 32-bit LSB executable, Intel 80386        &    	6742      &    	1.15    \\
				ia64         & 3000          & 64       & Little     & ELF 64-bit LSB executable, IA-64              &    	5046      &    	2.75    \\
				m68k         & 3000          & 32       & Big        & ELF 32-bit MSB executable, Motorola 68020     &    	4440      &    	1.17    \\
				mips $\dagger$         & 2997          & 32       & Big        & ELF 32-bit MSB executable, MIPS, MIPS-II ($\dagger$ upsampled)      &    	418      &    	1.00    \\
				mips64el $\ddagger$     & 2998          & 64       & Little     & ELF 64-bit LSB executable, MIPS, MIPS64 rel2 ($\ddagger$ new)     &    	6430      &    	2.61    \\
				mipsel       & 2994          & 32       & Little     & ELF 32-bit LSB executable, MIPS, MIPS-II      &    	4396      &    	1.01    \\
				powerpc      & 2110          & 32       & Big        & ELF 32-bit MSB executable, PowerPC or cisco 4500 &    	3672      &    	1.29    \\
				powerpcspe $\ddagger$   & 3000          & 32       & Big        & ELF 32-bit MSB executable, PowerPC or cisco 4500 ($\ddagger$ new) &    	3976      &    	1.63    \\
				ppc64        & 2552          & 64       & Big        & ELF 64-bit MSB executable, 64-bit PowerPC or cisco 7500  &    	2900      &    	1.75    \\
				ppc64el $\ddagger$      & 2997          & 64       & Little     & ELF 64-bit LSB executable, 64-bit PowerPC or cisco 7500 ($\ddagger$ new)  &    	4370      &    	1.03    \\
				riscv64 $\ddagger$      & 3000          & 64       & Little     & ELF 64-bit LSB executable ($\ddagger$ new)                                &    	4513      &    	1.18    \\
				s390         & 2997          & 32       & Big        & ELF 32-bit MSB executable, IBM S/390                     &    	5562      &    	0.61    \\
				s390x        & 2994          & 64       & Big        & ELF 64-bit MSB executable, IBM S/390                     &    	4169      &    	1.03    \\
				sh4          & 3000          & 32       & Little     & ELF 32-bit LSB executable, Renesas SH                    &    	6003      &    	1.30    \\
				sparc        & 2997          & 32       & Big        & ELF 32-bit MSB executable, SPARC32PLUS, V8+ Required     &    	6111      &    	0.62    \\
				sparc64      & 2676          & 64       & Big        & ELF 64-bit MSB executable, SPARC V9, relaxed memory ordering  &    	3338      &    	1.37    \\
				x32 $\ddagger$          & 3000          & 32       & Little     & ELF 32-bit LSB executable, x86-64 ($\ddagger$ new)                        &    	4261      &    	1.56    \\

				\bottomrule
				Total          & \SamplesAllExaCount{}          &   --     &    --   & --  &           \ELFExaCount{}    &   \ELFExaSizeGB{}  \\ 
				\bottomrule
			\end{tabular}%
		}
	\end{table*}
%\end{comment}

Our dataset covers \ArchsExaCount{} distinct architectures, 
which is inline with and comparable to 
Clemens~\cite{clemens2015automatic} (\ArchsExaCountClemens{} architectures) and 
De Nicolao et al.~\cite{de2018elisa} (\ArchsExaCountDeNicolao{} architectures). 
Most of the CPU architectures overlap with existing works, but there 
are few new ones (as marked in Table~\ref{table:supported_architectures}). 
At the same time, the sample-sets used in our experiments have 
some significant differences compared to the state of the art. 
First, the total number of \SamplesAllExaCount{} samples in our experiments is several times larger than those used by both 
Clemens~\cite{clemens2015automatic} (\SamplesAllExaCountClemens{} samples) and 
De Nicolao et al.~\cite{de2018elisa} (\SamplesAllExaCountDeNicolao{} samples). 
Second, compared to existing works, our sample-set size per architecture is both 
larger and more balanced. 
Using more balanced sample-sets should give more accurate results when evaluating 
and comparing classifier performance, as imbalance of classes in the 
dataset can cause sub-optimal classification 
performance~\cite{japkowicz2003class}. 
When creating our sample-sets for each architecture, we had set forth 
several constraints. On the one hand, we decided that the minimum 
code section in the ELF file should be 4000 bytes (4K), as this is 
the code size where all classifiers were shown to converge and
provide high-accuracy at the same time 
(see Fig.2 in Clemens~\cite{clemens2015automatic}). 
On the other hand, we wanted to have the sample-sets as balanced 
as possible between all the architectures. 
Given these parameters, from the initial \ELFExaCount{} ELF files 
in the download dataset, our toolset filtered \SamplesAllExaCount{} 
samples with an approximate average of \ArchSetSizeAvgApprCount{} 
samples per architecture sample-set 
(Table~\ref{table:supported_architectures}). 
Importantly, our toolset can be parametrized to 
download more files, and to filter the sample-sets based on 
different criteria as dictated by various use cases. 
%Only code sections over 4000 bytes were selected to eliminate the negative effect of too small sample, even though \cite{clemens2015automatic} opted to not use this method. He acknowledged it would avoid many of the features having close to zero matches, but argued it is a more realistic scenario in the field. Regardless, only large enough samples were used to train the classifiers in this thesis as there was enough data to do so and because as seen in \cite{clemens2015automatic} results, all classifiers are near 90\% accuracy at that point, so it removes one random variable when evaluating the classifiers.

\subsubsection{Discussion}
An important point to clarify at this stage is the reasoning and the 
implications of selecting the Debian repositories as the source of code 
binaries, and the subsequent experimentation limited to ELF binaries. 
To the best of our knowledge Debian package repositories are 
the only at this time that provide a years long list of compiled binaries 
for such an extensive list of CPU and hardware architectures. Certainly 
other repositories (e.g., Fedora, Ubuntu, Raspbian) can also provide a good 
source of compiled binaries, but in our experience and opinion they would only 
be able to marginally improve the quality and the quantity of the 
ones provided by the Debian repository. This may also be the reason why other 
state of the art works relied on Debian repositories as well~\cite{clemens2015automatic,de2018elisa}. 
As a result of using the Debian packages as source of code binaries, this 
inherently limited the datasets and the experiments (both ours and ones in the 
state of the art~\cite{clemens2015automatic,de2018elisa}) to using only ELF format 
binaries. This however is not limiting or impacting the experiments 
or applicability of the methods, since we mainly work with the raw machine code 
(i.e., op-codes) extracted solely from the code sections. 
Equivalent raw machine code can be extracted in similarly easy ways from 
DOS MZ, COFF, PE32(+), MACH-O, BFLT, and virtually any other executable 
binary format. The only important condition is that, 
regardless of the binary file format those op-codes come from, they must be 
extracted from code sections which is a way to guarantee those op-code byte 
sequences represent valid instructions for the CPU architecture they represent 
(and which is specified in the binary format headers). 
Unfortunately, we are unaware of any substantial collection of non-ELF 
binaries that would cover an extensive list of CPU architectures. 
For example, despite the fact the PE32 format supports well x86, x86-64, x64, 
and added recently support for ARM thanks to Windows IoT 
for Raspberry Pi (which has an ARM CPU), there is not much support for 
other CPUs beyond that. 
To summarize, it is important to emphasize that the methods evaluated or 
presented in this paper are not limited only to ELF files, though we 
performed our experiments on code sections extracted solely from ELF files.

\subsection{Machine Learning}

We then continued with the experiment reconstruction and cross-validation 
of the state of the art.
For training and testing our machine learning classifiers, we used the 
following complete feature-set which consists of 293 features as follows. 
The first 256 features are mapped to \emph{Byte Frequency Distribution (BFD)} 
(used by Clemens~\cite{clemens2015automatic}). The next 4 features map to 
\emph{``4 endianness signatures''} (used by Clemens~\cite{clemens2015automatic}). 
The following 31 features are mapped to \emph{function epilog and function 
prolog signatures} for amd64, arm, armel, mips32, powerpc, powerpc64, 
s390x, x86 (developed by \texttt{angr} framework~\cite{shoshitaishvili2016state} 
and also used by De Nicolao et al.~\cite{de2018elisa}). The final 2 features map 
to \emph{``powerpcspe signatures''} that were developed specifically for 
this paper.~\footnote{These signatures are in the process of being contributed 
back to open-source projects such as \texttt{angr, binwalk}.}
To this end, we extract the mentioned features from the code sections 
of the ELF binaries in the sample-sets (column \emph{Sample-sets size in experiment} in Table~\ref{table:supported_architectures}). 
We then save the extracted features into a CSV file ready 
for input to and processing by machine learning frameworks. 
%The full binary is not used, because data sections can alter the byte frequencies or include byte sequences that match a signature for another architecture, which can lead into false classifications \cite{de2018elisa}. The possible negative effect of using full binary for training is tested as a part of this thesis.

In order to replicate and validate the approach of 
Clemens~\cite{clemens2015automatic}, we used the Weka framework along 
with exact list and settings of classifiers as used by the 
author (Table~\ref{table:clemens_classifiers}). 
%We used default values for most of the classifiers in Weka, as our experiment's iterations have shown the tweaking of the parameters did not have a significant effect on most of the classifiers~\cite{amancio2014systematic}. 
We used non-default parameters (acquired by manual tuning) only 
for neural net when training the classifier on our complete dataset, 
because the parameters used by Clemens~\cite{clemens2015automatic} were 
specific to their dataset. We also used only the list of architectures 
and features used by the author. 

	\begin{table*}%[b]{\linewidth}
		\captionof{table}{Classifiers and their settings as suggested by Clemens~\cite{clemens2015automatic}.}
		\label{table:clemens_classifiers}
		\resizebox{\textwidth}{!}{%
			\begin{tabular}{@{}lll@{}}
				\toprule
				Model               & Weka name            & Parameters                                                                                                                                                                     \\ 
				\midrule
				1 nearest neighbor (1-NN)                & IBk                  & -K 1 -W 0 -A "weka.core.neighboursearch.-LinearNNSearch -A "weka.core.-EuclideanDistance -R first-last""                                                                       \\
				3 Nearest neighbors (3-NN)                & IBk                  & -K 3 -W 0 -A "weka.core.neighboursearch.-LinearNNSearch -A "weka.core.-EuclideanDistance -R first-last""                                                                       \\
				Decision tree       & J48                  & -C 0.25 -M 2                                                                                                                                                                   \\
				Random tree         & RandomTree           & -K 0 -M 1.0 -V 0.001 -S 1                                                                                                                                                      \\
				Random forest       & RandomForest         & -I 100 -K 0 -S 1 -num-slots 1                                                                                                                                                  \\
				Naive Bayes         & NaiveBayes           & N/A                                                                                                                                                                            \\
				BayesNet            & Bayesnet             & \begin{tabular}[c]{@{}l@{}}-D -Q weka.classifiers.bayes.net.-search.local.K2 -- -P 1\\ -S BAYES -E weka.classifiers.bayes.net.-estimate.SimpleEstimator -- -A 0.5\end{tabular} \\
				SVM (SMO)           & SMO                  & \begin{tabular}[c]{@{}l@{}}-C 1.0 -L 0.001 -P 1.0E-12 -N 0 -V -1 -W 1 -K "weka.classifiers.functions.-supportVector.PolyKernel\\ -E 1.0 -C 250007"\end{tabular}                \\

				Logistic regression & SimpleLogistic       & -I 0 -M 500 -H 50 -W 0.0                                                                                                                                                       \\
				Neural net          & MultilayerPerceptron & -L 0.3 -M 0.2 -N 100 -V 0 -S 0 -E 20 -H 66                                                                                                                                     \\ 
				\midrule
				Neural net ($\dagger$ this paper)    & MultilayerPerceptron & -L 0.3 -M 0.2 -N 100 -V 20 -S 0 -E 20 -H 66 -C -I -num-decimal-places 10                                                                                                                                     \\ 
				\midrule
			\end{tabular}%
		}	
	\end{table*}

In order to replicate and validate the approach of De Nicolao et al.~\cite{de2018elisa}, 
we used scikit-learn~\cite{scikit-learn}. The authors used only logistic 
regression classifier to which they add L1 regularization as compared to 
Clemens~\cite{clemens2015automatic}. In this paper, we implemented 
the logistic regression classifier both in scikit-learn and 
Keras~\cite{chollet2015keras} (a high-level API for neural networks) in 
order to see if the framework used has any effect on the classification 
accuracy. We also used only the list of architectures and features used by the authors.

\subsection{Hardware and Software}

To perform the work from this paper we used multiple combinations of software, 
as summarized in Table~\ref{table:software_hardware}. 
\begin{comment}
%
For the dataset acquisition workflow we used bash version 4.2.46(2), 
python version 3.6.6, and wget version 1.14.
%
To run the machine learning experiments we used Keras version 2.2.4, 
scikit-learn version 0.20.0, and Weka version 3.8.3. 
Both scikit-learn and Keras used Tensorflow as the back end. 
%
For the OpenAPI web-services we used python version 3.5.2, 
and flask-restplus version 0.12.1.
%
Radare2 plugin employed Radare2 version 3.3.0, and 
python version 3.5.3.
\end{comment}
%
In terms of hardware we used two main machines. 
The server used for \emph{dataset collection, and data pre- and post-processing} 
has a 4-cores CPU Intel(R) Xeon(R) E7-8837  \@ 2.67GHz with 16 GB 
of DDR3, and was running CentOS 7.4 with the kernel 3.10.0-957.5.1.el7.x86\_64. 
The host used for \emph{machine learning tasks} is a standard PC with a 
6-core CPU Intel(R) Core(TM) i7-8700k 5.00GHz CPU with 12 threads, 
GTX 970 graphics card, 16 GB of DDR4 and was running Windows 10. 

	\begin{table}%[b]{\linewidth}
		\captionof{table}{Software configurations used at various stages of our experiments.}
		\label{table:software_hardware}
			\begin{tabular}{@{}llc@{}}
				\toprule
                Type        &       Role        &       Configuration       \\ 
				\midrule
                Software    &       Dataset acquisition     &       \specialcell{bash 4.2.46(2) \\ python 3.6.6 \\ wget 1.14}       \\
                \midrule
                Software    &       ML experiments     &       \specialcell{Keras 2.2.4 (TF) \\ scikit-learn 0.20.0 (TF) \\ Weka 3.8.3}       \\
                \midrule
                Software    &       OpenAPI web-services    &       \specialcell{python 3.5.2 \\ flask-restplus 0.12.1}       \\
                \midrule
                Software    &       Radare2 plugin    &       \specialcell{Radare2 3.3.0 \\ python 3.5.3}       \\
                \bottomrule
			\end{tabular}%
	\end{table}

%%%%%%%%%%%%%%%%%%%%%%%%%%%%%%%%%%%%%%%%%%%%%%%%%%%%%%%%%%%%%%%%%%%%%%%%%%%%%%%%

\section{Results and analysis}
\label{sec:results}

\begin{comment}
In this section we go through the execution and the results of the experimental 
evaluation and cross-validation of existing work. We test the classifiers' 
performance and feature-sets in multiple different situations. Such scenarios 
include training and testing with code-only sections, testing with different 
sample sizes as well as testing and training+testing the performance of the 
classifiers on complete binaries. We also highlight our novel contributions 
and the results of their application. 
%Also, classifiers are trained and tested with full binaries to see if it has an effect on classifier performance.
\end{comment}

\subsection{Classifier performance when training and testing with code-only sections}

First, we compare the performance of multiple classifiers trained on 
code-only sections, when classifying code-only input. For this we use 
10-fold cross validation and the features extracted from code-only 
sections of the test binaries. Also, we evaluate the effect of various 
feature-sets on classification performance by calculating performance 
measures with the ``all features'' set, BFD-only features-set, 
and BFD+endianness features-set. We then cross-validate the results 
by Clemens~\cite{clemens2015automatic} as well as compare them to our 
results. 

Using Weka frameworks and the settings presented in Table~\ref{table:clemens_classifiers}, 
we trained and tested multiple different classifiers using different 
feature-sets. BFD corresponds to using only byte frequency 
distribution, while BFD+endianness adds the architecture endianness 
signatures introduced by Clemens~\cite{clemens2015automatic}. 
The complete data set includes the new architectures as well as the 
new signatures for \texttt{powerpcspe} (see also Section~\ref{sec:powerpcspe}). 
The performance metrics are weighted averages, i.e., sum of the metric 
through all the classes, weighted by the number of instances in the 
specific architecture class. The results are also compared to the 
results presented by~\cite{clemens2015automatic}, and can be observed in 
Table~\ref{table:classifier_performance}. We marked with asterisk (*) 
the results that we obtained using different parameters than those 
in~\cite{clemens2015automatic}. 
	
	\begin{table*}%[b]{\linewidth}
		\captionof{table}{Classifier performance with different feature-sets. For comparison, we also present results from Clemens~\cite{clemens2015automatic}.}
		\label{table:classifier_performance}
			\begin{tabular}{@{}lccccccccc@{}}
				\toprule
				Classifier          & Precision & Recall & AUC   & \specialcell{F1 \\ measure} & \specialcell{Accuracy \\ All features \\ (293 features)} & \specialcell{Accuracy \\ BFD+endian} & \specialcell{Accuracy \\ BFD+endian \\ Clemens~\cite{clemens2015automatic}}  & \specialcell{Accuracy \\ BFD}   & \specialcell{Accuracy \\ BFD \\ Clemens~\cite{clemens2015automatic}} 					 \\ \midrule
				1-NN                & 0.983     & 0.983  & 0.991 & 0.983      & 0.983    & 0.911      & 0.927              & 0.895 & 0.893       \\
				3-NN                & 0.994     & 0.994  & 0.999 & 0.994      & 0.993    & 0.957      & 0.949              & 0.902 & 0.898       \\
				Decision tree       & 0.992     & 0.992  & 0.998 & 0.992      & 0.992    & 0.993      & 0.980              & 0.936 & 0.932       \\
				Random tree         & 0.966     & 0.966  & 0.982 & 0.966      & 0.965    & 0.953      & 0.929              & 0.899 & 0.878       \\
				Random forest       & 0.996     & 0.996  & 1.000 & 0.996      & 0.996    & 0.992      & 0.964              & 0.904 & 0.904       \\
				Naive Bayes         & 0.991     & 0.991  & 0.999 & 0.991      & 0.990    & 0.990      & 0.958              & 0.932 & 0.925       \\
				BayesNet            & 0.992     & 0.992  & 1.000 & 0.992      & 0.991    & 0.994      & 0.922              & 0.917 & 0.895       \\
				SVM (SMO)           & 0.997     & 0.997  & 1.000 & 0.997      & \textbf{0.997}    & \textbf{0.997}      & \textbf{0.983}              & 0.931 & 0.927       \\
				Logistic regression & 0.989     & 0.988  & 0.998 & 0.989      & 0.988    & \textbf{0.997}      & 0.979              & 0.939 & 0.930       \\
				Neural net          & 0.995*    & 0.994* & 1.000*& 0.994*     & 0.994*   & 0.919      & 0.979              & \textbf{0.940} & \textbf{0.940}       \\
				\bottomrule
			\end{tabular}%
	\end{table*}
	
As can be seen in Table~\ref{table:classifier_performance}, the 
results are inline with the ones presented by Clemens~\cite{clemens2015automatic}, 
even though we constructed and used our own datasets. 
%The largest differences to the results presented by \cite{clemens2015automatic} were BayesNet's sensibly higher accuracy of 99.4\% compared to the accuracy reported by \cite{clemens2015automatic} of 92.2\% and neural net's sensibly lower accuracy of 91.9\% compared to 97.9\% when using both BFD and endianness features. 
%
In our experiments, the complete data set (with added 
architectures and all features considered) increased the accuracy 
of all classifiers (in some cases by up to 7\%) when compared to 
the results of Clemens~\cite{clemens2015automatic}. 
This could be due to a combination of larger overall dataset, 
more balanced sets for each CPU architecture class, and the use of only binaries that have code 
sections larger than 4000 bytes.

\subsubsection{Effect of test sample code size on classification performance}

\begin{figure}[htb]
		\includegraphics[keepaspectratio,width=\linewidth]{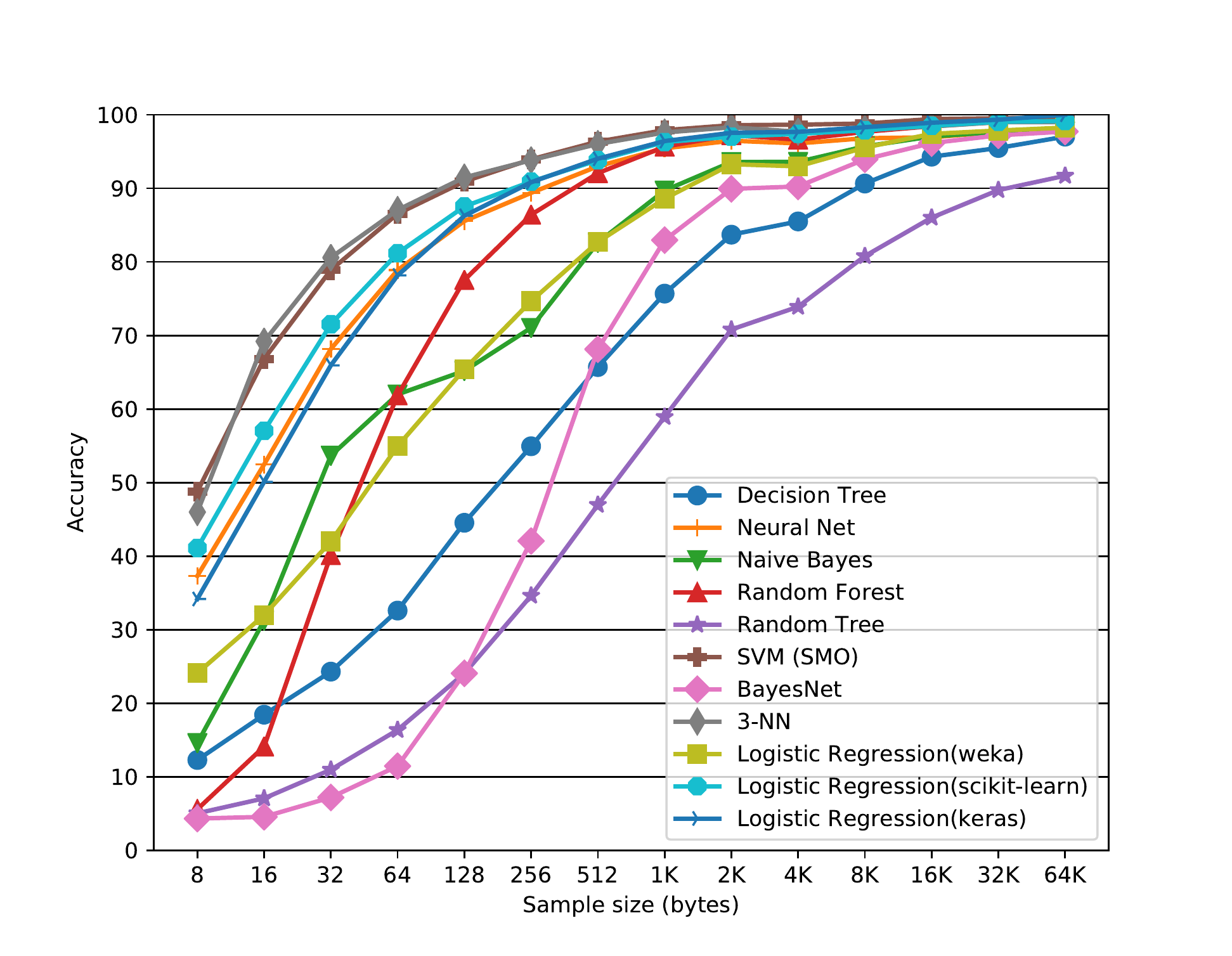}
		\captionof{figure}{Impact of the test sample size on classifier performance.}
		\label{fig:accuracy_vs_sample_size}
	\end{figure}

Next, we study if the sample size has an effect on the 
classification performance. For this, we test the classifiers 
against a test set of code sections with increasingly varying size, 
as also performed by both Clemens~\cite{clemens2015automatic} and 
De Nicolao et al.~\cite{de2018elisa}. 
If the performance of such classifiers is good enough with only 
small fragments of the binary code, those classifiers could be 
used in environments where only a part of the executable is present. 
For example, small (128 bytes or less) code size fragments could be 
encountered in digital forensics when only a portion of malware or 
exploit code is successfully extracted from an exploited smartphone 
or IoT device. 
For this test, the code fragments were taken from code-only sections 
using random sampling in order to avoid any bias that could come from 
using only code from the beginning of code sections~\cite{clemens2015automatic}. 
%On the other hand, this will decrease reliability as acquiring exactly the same data is unlikely to happen.
%
%To see the effect of the size of the classified binary, samples of code sections of varying sizes were taken and the classifiers trained using the whole code sections were tested against them. Small (128 bytes or less) sample code sizes could be encountered in digital forensics, for example when only a portion of malware or exploit code is successfully extracted from a smartphone or an IoT device. 
%Bigger samples (i.e., thousands of bytes) could already be full binaries encountered in IoT device firmware images. 
We present the results of this test in Figure~\ref{fig:accuracy_vs_sample_size}. 
	
When testing with varying size of the test input, SVM performed the best with 
almost 50\% accuracy even with the smallest sample size of 8 bytes. Also, SVM 
along with 3 nearest neighbors achieved 90\% accuracy at 128 bytes. 
Logistic regression implemented in scikit-learn and Keras were very close 
performance-wise, both implementations achieving 90\% accuracy at 256 bytes. 
Surprisingly, logistic regression implemented in Weka under-performed and 
required 2048 bytes to reach 90\% accuracy. 
Cross-validating and comparing the result, the classifiers that performed 
the best in our varying sample size experiments also performed well in the 
experiments by Clemens~\cite{clemens2015automatic}. 
On the other hand, in this experiments not all the classifiers achieved 
90\% accuracy at 4000 bytes as experienced by Clemens~\cite{clemens2015automatic}. 
For example, at 4000 bytes the Decision Tree and Random Tree classifiers in 
our case achieved accuracy of only 85\% and 75\% respectively. 

%%%%%%%%%%%%%%%%%%%%%%%%%%%%%%%%%%%%%%%%%%%%%%%%%%%%%%%%%%%%%%%%%%%%%%%%%%%%%%%%

\subsubsection{Effect of different frameworks on performance of logistic regression}

From all the different classifiers available, De Nicolao et al.~\cite{de2018elisa} 
used logistic regression only and used the scikit-learn as their machine learning 
framework of choice. Logistic regression has a couple of parameters that affect the 
classification performance. The authors used \emph{grid search} to identify the 
best value for \emph{C}, which stands for inverse of regularization 
strength and found the value of 10000 to give the best results in their case. 
Since the dataset itself affects the result, we ran \emph{grid search} for the 
scikit-learn model developed based on our dataset. The \emph{C} values of 
10000, 1000, 100, 10, 1, 0.1 were tested and we found that for our case the value 
of 1000 gave the best results. Similarly, for the Keras model, we found for \emph{C} 
the value of 0.0000001 to provide the best accuracy. The Table~\ref{table:logistic_regression} 
present our results of logistic regression using 10-fold cross-validation on 
code-only sections when tested in all different frameworks. 

With this experiment, for example, we found that for the same dataset, the 
logistic regression implemented in scikit-learn and Keras provided better results 
(F1 measures of 0.998) when compared to Weka (F1 measures of 0.989). 
%Scikit-learn and Keras give similar results except Keras giving 99.7\% accuracy compared to scikit-learn's 99.6\%.

	\begin{table}[htb]%{\linewidth}
		\captionof{table}{Logistic regression 10-fold cross validation performance in different machine learning frameworks.}
		\label{table:logistic_regression}
		\begin{tabular}{@{}lccccc@{}}
			\toprule
			Classifier          			& Precision & Recall & AUC   & F1 measure & Accuracy \\ \midrule
			Weka                			& 0.989     & 0.988  & 0.998 & 0.989      & 0.988    \\
			scikit-learn                	& 0.998     & 0.998  & 0.998 & 0.998      & 0.996    \\
			Keras			       			& 0.998     & 0.998  & 0.998 & 0.998      & \textbf{0.997}    \\ \bottomrule
		\end{tabular}
	\end{table}

\subsection{Classifier performance when training with code-only sections and testing with complete binaries}

%To see how well classifiers perform classifying full binaries, the classifiers trained with only code sections of binaries are tested against a separate test set consisting of 500 full binaries per architecture. The addition of data sections in the full binaries will have an effect on the byte frequency distribution, but according to \cite{de2018elisa}, it did not have an effect on classification accuracy. In real life case, it might not always be possible to extract code sections from binaries, so this could reflect on how well the classifiers would perform in real life situations. 
We also explored how well the classifiers perform when given the task to 
classify a complete binary (i.e., containing headers, and code and data sections). 
In fact, De Nicolao et al.~\cite{de2018elisa} tested their classifier performance on 
complete binaries (i.e., full executables). 
Therefore, in this work we test all the different classifiers used by 
Clemens~\cite{clemens2015automatic} and De Nicolao et al.~\cite{de2018elisa} against 
complete binaries using a separate test set consisting of 500 binaries for each 
architecture (which is about 1.5 times more than in~\cite{de2018elisa}). 
The classifiers we used in this test are still previously trained using 
code-only sections %while in Section \ref{section:train_full_binaries} the classifiers are also trained using the full binaries. 
The results of our tests for this experiments can be seen in 
Table~\ref{table:test_on_full_binaries}. 

	\begin{table*}%{\linewidth}
		\captionof{table}{Performance of classifiers (in our Weka, unless specified otherwise) when trained on code-only sections and testing on complete binaries.}
		\label{table:test_on_full_binaries}
		\begin{tabular}{@{}lccccc@{}}
			\toprule
			Classifier          			& Precision & Recall & AUC   & F1 measure & Accuracy \\ \midrule
			1-NN                			& 0.871     & 0.742  & 0.867 & 0.772      & 0.741    \\
			3-NN                			& 0.876     & 0.749  & 0.892 & 0.773      & 0.749    \\
			Decision tree       			& 0.845     & 0.717  & 0.865 & 0.733      & 0.716    \\
			Random tree         			& 0.679     & 0.613  & 0.798 & 0.619      & 0.613    \\
			Random forest       			& 0.912     & \textbf{0.902}  & \textbf{0.995} & \textbf{0.892}      & \textbf{0.901}    \\
			Naive Bayes        			 	& 0.807     & 0.420  & 0.727 & 0.419      & 0.420    \\
			Bayes net          				& 0.886     & 0.844  & 0.987 & 0.840      & 0.844    \\
			SVM (SMO)           			& 0.883     & 0.733  & 0.971 & 0.766      & 0.732    \\
			Logistic regression (Weka) 		& 0.875     & 0.718  & 0.978 & 0.728      & 0.718    \\
			Logistic regression (scikit-learn) 	& 0.913 	& 0.780  & 0.780 & 0.794	  & 0.579    \\
			Logistic regression (Keras)     & \textbf{0.921}     & 0.831  & 0.831 & 0.839      & 0.676    \\
			Neural net         				& 0.841     & 0.452  & 0.875 & 0.515      & 0.451    \\ 
			%Average (our)     				& 0.859     & 0.708  & 0.881 & 0.724      & 0.678    \\ 
			\midrule
			Average (De Nicolao et al.~\cite{de2018elisa}) 		& 0.996		& 0.996  & 0.998 & 0.996	  & -		 \\
			\bottomrule
		\end{tabular}
	\end{table*}

Our analysis shows that Random Forest performed the best by having the highest 
performance measures of 0.901 for accuracy and 0.995 for AUC. The logistic 
regression implemented in scikit-learn did not perform as well as experiences 
by De Nicolao et al.~\cite{de2018elisa}. %with reported averages of 0.99 in all performance measures except accuracy that was not presented. 
The time to classify all the binaries in this test set took only a couple of 
seconds on all algorithms except the Nearest Neighbor algorithms which 
took approximately 15 minutes. One of the reasons for this is because 
the Nearest Neighbor algorithm is a lazy classifier, and the model is only 
built when data needs to be classified. %, as explained in Section \ref{subsection:lazy_classifiers}. 
	
\begin{figure}
		\includegraphics[keepaspectratio,width=\linewidth]{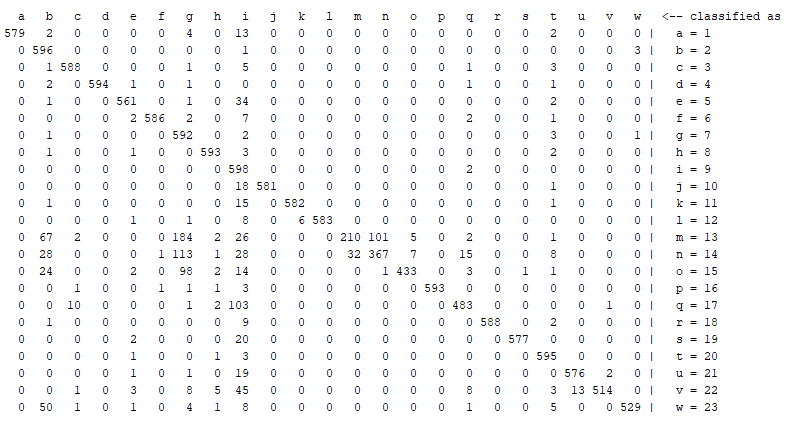}
		\captionof{figure}{Confusion matrix of Random Forest classification results when testing complete binaries. Columns show the predicted class frequencies while rows show the true class frequencies.}
		\label{fig:random_forest_confusion_matrix}
	\end{figure}

We also present in Figure~\ref{fig:random_forest_confusion_matrix} the 
confusion matrix of the best performing classifier in this test, 
i.e., the Random Forest classifier . The columns represent the class 
frequencies \emph{predicted} by the model while the rows present \emph{true} 
class frequencies. Everything off from the diagonal is a misclassification. 
The alphabets represent the \ArchsExaCount{} architecture classes in the alphabetical order as presented 
in Table~\ref{table:supported_architectures}. 
For example, looking at the confusion matrix it is possible to see that the 
\texttt{i386 (g)} and the \texttt{m68k (i)} architectures caused over 
70\% of the misclassifications. Therefore, one direction in future work is 
to find the root cause of this, and develop better discriminating signature 
for these architectures. 

%%%%%%%%%%%%%%%%%%%%%%%%%%%%%%%%%%%%%%%%%%%%%%%%%%%%%%%%%%%%%%%%%%%%%%%%%%%%%%%%

%\subsubsection{Effect of classifier implementation on its performance}
\subsubsection{Testing Machine Learning as cloud implementations}

\begin{comment}
\fixme{
	To test if the classifier implementation has an effect, random forest was also trained in Azure Machine Learning platform. The used parameters were "Number of decision trees: 100, Maximum depth of the decision trees: 32, Number of random splits per node: 128, Minimum number of samples per leaf node: 1". The classifier was trained using the whole data set, and tested against the full binary test set. Also, decision jungle was tested as it is closely related to random forest as explained in Section \ref{subsection:trees}. The parameters used to train decision jungle were: "Number of decision DAGs: 100, Maximum depth of the decision DAGs: 32, Maximum width of the decision DAGs: 128, Number of optimization steps per decision DAG layer: 2048". All the parameters were empirically tested and selected. Increasing the number of decision trees or DAGs increased the accuracy to some point. Other research has seen same kind of behavior with random forest \cite{shafieian2015cloudzombie}. The results can be seen in Table \ref{table:azure_results}. Results from Weka were added to make it easier to compare the results.
}
\end{comment}

In addition to the existing approaches and tools detailed above, we also 
employed Azure Machine Learning platform to test code-only trained 
classifiers with complete binaries as input. 
%Azure platform provides several classifiers for both multi-class 
%classifications as well as binary ones (which can also be used in 
%multi-class cases with ``One-vs-All'' plugin provided by Azure). 
%The classifiers will be chosen based on what classifiers will perform the best on other frameworks. 
%
An example of our setup and workflow in Azure platform is 
presented in Figure~\ref{fig:azure}, and its (comparative) performance 
is presented in Table~\ref{table:azure_results}. 
%
%In the example, the classifier is trained on all of the training data and tested against a separate test set containing 500 full binaries for each architecture.
%
In summary, the Random Forest classifier implemented in Azure performed 
sensibly better when compared to the one implemented in Weka. 
This once again highlights the long-standing research challenge that, 
even when using the same methods or algorithms, the implementations do matter 
and can affect the research results whether in computer science~\cite{kriegel2017black} 
or other scientific fields~\cite{durlak2008implementation}. 
%Random Forest in Azure had over 6 percentage units better accuracy, precision and recall than Weka implementation.

	\begin{table}[htb]%{\linewidth}
		\captionof{table}{Performance of Random Forest and Decision Jungle classifiers trained and tested in Azure Machine Learning platform.% For comparison, performance results for Random Forest implemented in Weka were copied from Table~\ref{table:test_on_full_binaries}.
		}
		\label{table:azure_results}
		\begin{tabular}{@{}lccc@{}}
			\toprule
			Classifier      		& Recall 	& Precision & Accuracy \\ 
			\midrule
			Decision jungle (Azure)	& 0.964    	& 0.964    	& 0.964  \\ 
			Random forest (Azure)	& 0.974    	& 0.974    	& \textbf{0.974}  \\ 
			\midrule
			\specialcell{Random forest (Weka) \\ (Table~\ref{table:test_on_full_binaries}, comparison)} 	& 0.902   	& 0.912 	& 0.901  \\ 
			\bottomrule
		\end{tabular}
	\end{table}

%%%%%%%%%%%%%%%%%%%%%%%%%%%%%%%%%%%%%%%%%%%%%%%%%%%%%%%%%%%%%%%%%%%%%%%%%%%%%%%%

\subsection{Classifier performance when training and testing with complete binaries}
\label{section:train_full_binaries}

%Finally, the classifiers are trained using full binaries to see if they perform better or worse compared to classifiers trained with only code sections when classifying full binaries.

We then verify how the training of classifiers using complete binaries 
(i.e., not code-only sections as in previous works) affects their 
performance when given complete binaries as test input. 
For this, we selected some of the best performing classifiers from the 
previous experiments. 
Then we trained those classifiers with a training set consisting of 
1000 complete binaries for each of those \ArchsExaCount{} architectures. 
Finally, we tested them against a test set of 1000 full binaries for 
each architecture. The results of this experiment can be seen in 
Table~\ref{table:full_binary_training}. For comparison, along with 
the performance of classifiers trained and tested on complete binaries, 
we also present the performance of exact same classifiers when tested 
in code-only experiments above. 

	\begin{table*}%{\linewidth}
		\captionof{table}{Performance on complete binaries for classifiers trained with complete binaries. Random Forest and Random Jungle are missing AUC and F1 measures as Azure platform does not output them.}
		\label{table:full_binary_training}
			\begin{tabular}{@{}lcccccc@{}}
				\toprule
				Classifier                   & Precision & Recall & AUC   & F1 measure & Accuracy & \specialcell{Accuracies \\ (Tables \ref{table:test_on_full_binaries} and \ref{table:azure_results})} \\ \midrule
				Random forest (Weka)         & 0.987     & 0.987  & 1.000 & 0.987      & 0.986    & 0.901                    \\
				Random forest (Azure)        & 0.992     & 0.992  & -     & -          & \textbf{0.992}    & 0.974                    \\
				Decision jungle (Azure)      & 0.989     & 0.989  & -     & -          & 0.989    & 0.964                    \\
				Logistic regression (Weka)   & 0.983     & 0.983  & 0.999 & 0.983      & 0.982    & 0.718                    \\
				Logistic regression (scikit) & 0.985     & 0.984  & 0.984 & 0.984      & 0.971    & 0.579                    \\
				Logistic regression (Keras)  & 0.990     & 0.989  & 0.989 & 0.989      & 0.980    & 0.676                    \\
				SVM (SMO)                    & 0.975     & 0.975  & 0.997 & 0.975      & 0.974    & 0.732                    \\
				\bottomrule
			\end{tabular}%
	\end{table*}

Previous work did not propose or use complete binaries for training the 
classifiers, and used complete binaries only as classification test 
inputs~\cite{clemens2015automatic,de2018elisa}. 
As can be seen in Table~\ref{table:full_binary_training}, all the 
classifiers we tested achieved over 97\% accuracy, while 
the Random Forest implemented in Azure platform performed the 
best with 99.2\% accuracy. 
%and the results look very good compared to the performance of classifiers trained with only code sections when classifying full binaries. All the tested classifiers achieved over 97\% accuracy with random forest implemented in Azure performed the best with 99.2\% accuracy. The accuracy increased compared to the results presented earlier, with logistic regression implemented in scikit-learn achieving over 40 percentage units better accuracy.
%
The proposal of using complete binaries for both training and testing 
the classifiers, as well as the experimental confirmation that the accuracy 
of classification is comparable to existing approaches and is very high 
(e.g., up to 99.2\%), is another incremental but novel contribution 
of this paper to the field. 

\subsubsection{Discussion}
Up to this point, we evaluated our ``complete binary'' training and recall only using 
ELF format binaries, and we presented the reasons for experimenting only with ELF 
format binaries in Section~\ref{sec:datasets}. 
We are aware that executable binary file formats such as PE32, COFF, MACH-O, 
and including ELF, even when compiled for the same CPU architecture, may certainly 
have major structural difference between each other, despite the fact that their code 
sections may contain similar, equivalent or highly-comparable op-code byte sequences. 
This is due to many factors, including but not limited to file format header variations, 
compiler and its options, the way the data (e.g., strings, constants, initial values for variables) 
is stored in the binary file and subsequently referenced within code sections. 
Such structural differences could in turn influence both the training and the recall 
of the ``complete binary'' method we introduced. 
Therefore, we plan as a future work to further evaluate and improve the ``complete binary'' 
approach for non-ELF-only cases such as when dealing with other homogeneous binary 
format datasets (e.g., only PE32) or with heterogeneous binary format datasets 
(e.g., combination of ELF, PE32, MACH-O).

\subsection{The special case of signatures and features for powerpcspe}
\label{sec:powerpcspe}

During the testing of classifiers performance, we observed that some 
architectures were the root cause behind the most false matches performed 
by the classifiers. For example, the binary code for \texttt{powerpc} (PPC)
and \texttt{powerpcspe} (PPCspe) are essentially the same, the only difference 
being the presence of SPE instructions~\footnote{SPE stands for 
Signal Processing Engine. SPE instructions perform floating-point 
operations on the integer registers.} in \texttt{powerpcspe} ISA. 
Therefore, we had to create custom signatures to be able to more accurately 
distinguish between the \texttt{powerpc} and \texttt{powerpcspe} code sequences. 
We created the signatures by comparing the instructions between the two 
architectures and finding unique ones that only appear in one of the architectures. 
%The two signatures composed of many floating point and integer operations that appeared only in powerpcspe architecture. The many different floating point operations were so similar in byte representation, that it was possible to create a single signature for them. The same applied to the integer operations. 
For each analyzed architecture, the dataset for this experiment consisted 
of 1000 complete binaries for training, and the same amount for testing. 
To run this test, we employed Random Forest classifier as implement in 
Weka framework, and the results are presented in Table~\ref{table:signature_significance}. 

	\begin{table}%[b]{\linewidth}
		\captionof{table}{F1 score with Random Forest trained and tested with complete binaries in Weka without and with powerpcspe signatures/features.}
		\label{table:signature_significance}
		\begin{tabular}{@{}l|rr|r@{}}
			\toprule
			Architecture & F1 score          &       & Improvement            \\ 
			\cmidrule(r){2-3}
			& without signature & with signature &  \\
			PPC      & 0.868             & 0.894 & 2.99\%       \\
			PPCspe   & 0.888             & 0.906 & 2.02\%       \\
			\bottomrule
		\end{tabular}
	\end{table}

Our analysis shows that the addition of the two additional \texttt{powerpcspe} 
signatures increased the F1 score of Random Forest implemented in 
Weka by about two percentage units. %which corresponds to about two to three percentages performance increase.
The confusion matrix for \texttt{powerpc} and \texttt{powerpcspe} architectures 
without the signatures is presented for comparison next to the confusion matrix 
when the two signatures/features were used, and can be seen in 
% BUG: Shows 4.4 (like subsection reference) instead of 8.
%Table~\ref{table:confmatrixspe}.
Table~$8$. 

%the additional signatures can be seen in Figure \ref{table:without_sig} and confusion matrix with the additional signatures can be seen in \ref{table:with_sig}.

	\begin{table}%[b]{\linewidth}
		\label{table:confmatrixspe}
		\captionof{table}{Confusion matrix of powerpc and powerpcspe architectures without and with the additional powerpcspe signatures/features.}
		\begin{tabular}{@{}l|cc|cc@{}}
			\toprule
			& \specialcell{Predicted \\ PPC } & \specialcell{(w/o signature) \\ PPCspe } & \specialcell{Predicted \\ PPC } & \specialcell{(with signature) \\ PPCspe}   \\ 
			\midrule
			\specialcell{Actual: \\ PPC }    & 802                & 200                   & 843                & 160                   \\
			\midrule
			\specialcell{Actual: \\ PPCspe } & 39                 & 963                   & 36                 & 967 \\ %965                   \\
		    \bottomrule
		\end{tabular}
	\end{table}

\begin{comment}
	\begin{table}%[b]{\linewidth}
		\centering
		\label{table:with_sig}
		\captionof{table}{Confusion matrix of powerpc and powerpcspe architectures with the additional signatures.}
		\begin{tabular}{@{}lll@{}}
			\toprule
			& Predicted: powerpc & Predicted: powerpcspe \\ \midrule
			Actual: powerpc    & 843                & 160                   \\
			Actual: powerpcspe & 36                 & 965                   \\ \bottomrule
		\end{tabular}
	\end{table}
\end{comment}

%The additional signature helped to to correctly classify about 40 more powerpc samples. The effect of the signatures can be also seen by running the same data set on a logistic regression classifier and inspecting the coefficients for powerpc and powerpcspe. They are presented in Figure \ref{fig:powerpcspe_logistic_regression}.
	
%The negative coefficients in powerpc (class 13) means the relationship between the class and the feature is negative, i.e., if the two signatures are found from a binary, it is less likely to be a powerpc binary. Looking at class 14 representing powerpcspe, one of the signatures correlates positively on it, meaning that if the signature is found from a binary, it is more likely to be a powerpcspe binary.

The development and addition of the \texttt{powerpcspe} discriminating 
features, and the experimental confirmation that they improve the 
overall classification accuracy and confusion matrix, is another 
contribution of this paper to the field. 

%%%%%%%%%%%%%%%%%%%%%%%%%%%%%%%%%%%%%%%%%%%%%%%%%%%%%%%%%%%%%%%%%%%%%%%%%%%%%%%%

\section{Conclusion}
\label{sec:conclusion}

In this paper we tried to bridge multiple gaps in the field of automated and 
precise identification of architecture and endianness of binary files 
and object code. 
For this, we developed from scratch the toolset and datasets that are lacking 
in this research space. As a result, we contribute a comprehensive collection 
of \emph{open data}, \emph{open source}, and \emph{open API} web-services. 
We performed experiment reconstruction and cross-validation of the 
effectiveness, efficiency, and results of the state of the art methods 
by Clemens~\cite{clemens2015automatic} and De Nicolao et al.~\cite{de2018elisa}. 
When training and testing classifiers using solely code-sections 
from compiled binary files (e.g. ELF binaries), all our classifiers 
performed equally well achieving over 98\% accuracy. 
We have shown that our results are generally inline with the state of art, 
and in some cases we managed to outperform previous work by up to 7\%. 
Additionally, we provided with this work novel contributions to the field. 
One contribution is the proposal and confirmation that complete binaries 
can be successfully used for both training and testing machine learning 
classifiers. In this direction, we demonstrate a 99.2\% accuracy using 
Random Forest classifiers implemented in Azure platform. 
Another contribution is the development and validation of new 
discriminating features for \texttt{powerpc} and \texttt{powerpcspe} 
architectures. 
Finally, our work provides an independent confirmation of the general 
validity and soundness of both existing and newly proposed algorithms, 
features, and approaches.

\subsection{Future work}
\label{sec:futurework}

There are several directions for future work, and we plan to initially 
focus on the following. 
First, we would like to continuously expand the datasets in terms 
of size, number of supported architectures, and quality. We plan 
to achieve this by using community submitted data subsets as well 
as by setting up our own multi-architecture object code building 
infrastructures. 
Second, we plan to use crowdsourcing in order to get community's 
expert knowledge that would continously increase the performance of 
machine learning classifiers. One way to achieve this is to 
allow user to confirm or correct automatically classified results. 
Third, we plan to expand the work on architecture discriminating signatures. 
One idea is to develop advanced methods (e.g., based on machine learning) 
to automatically create signatures for more accurate discrimination of 
code's CPU architecture, for example by using op-code specifics, and 
function epilogs and prologs. 

%%%%%%%%%%%%%%%%%%%%%%%%%%%%%%%%%%%%%%%%%%%%%%%%%%%%%%%%%%%%%%%%%%%%%%%%%%%%%%%%

\bibliographystyle{ACM-Reference-Format}
\bibliography{sample-base}


\begin{thebibliography}{56}

%%% the Web address.

\ifx \showCODEN    \undefined \def \showCODEN     #1{\unskip}     \fi
\ifx \showDOI      \undefined \def \showDOI       #1{#1}\fi
\ifx \showISBNx    \undefined \def \showISBNx     #1{\unskip}     \fi
\ifx \showISBNxiii \undefined \def \showISBNxiii  #1{\unskip}     \fi
\ifx \showISSN     \undefined \def \showISSN      #1{\unskip}     \fi
\ifx \showLCCN     \undefined \def \showLCCN      #1{\unskip}     \fi
\ifx \shownote     \undefined \def \shownote      #1{#1}          \fi
\ifx \showarticletitle \undefined \def \showarticletitle #1{#1}   \fi
\ifx \showURL      \undefined \def \showURL       {\relax}        \fi
% invisible to TeX
\providecommand\bibfield[2]{#2}
\providecommand\bibinfo[2]{#2}
\providecommand\natexlab[1]{#1}
\providecommand\showeprint[2][]{arXiv:#2}

\bibitem[\protect\citeauthoryear{??}{bin}{[n. d.]}]%
        {binwalk}
 \bibinfo{year}{[n. d.]}\natexlab{}.
\newblock \bibinfo{title}{{binwalk -- Firmware Analysis Tool}}.
\newblock
\newblock
\urldef\tempurl%
\url{https://github.com/binwalk/binwalk}
\showURL{%
\tempurl}


\bibitem[\protect\citeauthoryear{??}{noh}{[n. d.]}]%
        {nohr1993unix}
 \bibinfo{year}{[n. d.]}\natexlab{}.
\newblock \bibinfo{booktitle}{\emph{UNIX System V: understanding ELF object
  files and debugging tools}}.
\newblock


\bibitem[\protect\citeauthoryear{Beebe, Maddox, Liu, and Sun}{Beebe
  et~al\mbox{.}}{2013}]%
        {beebe2013sceadan}
\bibfield{author}{\bibinfo{person}{Nicole~L Beebe}, \bibinfo{person}{Laurence~A
  Maddox}, \bibinfo{person}{Lishu Liu}, {and} \bibinfo{person}{Minghe Sun}.}
  \bibinfo{year}{2013}\natexlab{}.
\newblock \showarticletitle{Sceadan: using concatenated n-gram vectors for
  improved file and data type classification}.
\newblock \bibinfo{journal}{\emph{IEEE Transactions on Information Forensics
  and Security}} \bibinfo{volume}{8}, \bibinfo{number}{9}
  (\bibinfo{year}{2013}), \bibinfo{pages}{1519--1530}.
\newblock


\bibitem[\protect\citeauthoryear{Blanco and Eissler}{Blanco and
  Eissler}{2012}]%
        {blanco2012one}
\bibfield{author}{\bibinfo{person}{Andr{\'e}s Blanco} {and}
  \bibinfo{person}{Matias Eissler}.} \bibinfo{year}{2012}\natexlab{}.
\newblock \bibinfo{title}{One firmware to monitor 'em all}.
\newblock
\newblock


\bibitem[\protect\citeauthoryear{Brumley, Jager, Avgerinos, and
  Schwartz}{Brumley et~al\mbox{.}}{2011}]%
        {brumley2011bap}
\bibfield{author}{\bibinfo{person}{David Brumley}, \bibinfo{person}{Ivan
  Jager}, \bibinfo{person}{Thanassis Avgerinos}, {and}
  \bibinfo{person}{Edward~J Schwartz}.} \bibinfo{year}{2011}\natexlab{}.
\newblock \showarticletitle{{BAP: A binary analysis platform}}. In
  \bibinfo{booktitle}{\emph{International Conference on Computer Aided
  Verification}}. Springer, \bibinfo{pages}{463--469}.
\newblock


\bibitem[\protect\citeauthoryear{Cha, Avgerinos, Rebert, and Brumley}{Cha
  et~al\mbox{.}}{2012}]%
        {cha2012unleashing}
\bibfield{author}{\bibinfo{person}{Sang~Kil Cha}, \bibinfo{person}{Thanassis
  Avgerinos}, \bibinfo{person}{Alexandre Rebert}, {and} \bibinfo{person}{David
  Brumley}.} \bibinfo{year}{2012}\natexlab{}.
\newblock \showarticletitle{Unleashing mayhem on binary code}. In
  \bibinfo{booktitle}{\emph{2012 IEEE Symposium on Security and Privacy}}.
  IEEE, \bibinfo{pages}{380--394}.
\newblock


\bibitem[\protect\citeauthoryear{Chipounov and Candea}{Chipounov and
  Candea}{2010}]%
        {chipounov2010reverse}
\bibfield{author}{\bibinfo{person}{Vitaly Chipounov} {and}
  \bibinfo{person}{George Candea}.} \bibinfo{year}{2010}\natexlab{}.
\newblock \showarticletitle{Reverse engineering of binary device drivers with
  RevNIC}. In \bibinfo{booktitle}{\emph{Proceedings of the 5th European
  conference on Computer systems}}. ACM, \bibinfo{pages}{167--180}.
\newblock


\bibitem[\protect\citeauthoryear{Chipounov, Kuznetsov, and Candea}{Chipounov
  et~al\mbox{.}}{2011}]%
        {chipounov2011s2e}
\bibfield{author}{\bibinfo{person}{Vitaly Chipounov},
  \bibinfo{person}{Volodymyr Kuznetsov}, {and} \bibinfo{person}{George
  Candea}.} \bibinfo{year}{2011}\natexlab{}.
\newblock \showarticletitle{S2E: A platform for in-vivo multi-path analysis of
  software systems}. In \bibinfo{booktitle}{\emph{ACM SIGARCH Computer
  Architecture News}}, Vol.~\bibinfo{volume}{39}. ACM,
  \bibinfo{pages}{265--278}.
\newblock


\bibitem[\protect\citeauthoryear{Chollet et~al\mbox{.}}{Chollet
  et~al\mbox{.}}{2015}]%
        {chollet2015keras}
\bibfield{author}{\bibinfo{person}{Fran\c{c}ois Chollet} {et~al\mbox{.}}}
  \bibinfo{year}{2015}\natexlab{}.
\newblock \bibinfo{title}{Keras}.
\newblock \bibinfo{howpublished}{\url{https://keras.io}}.
\newblock


\bibitem[\protect\citeauthoryear{Clemens}{Clemens}{2015}]%
        {clemens2015automatic}
\bibfield{author}{\bibinfo{person}{John Clemens}.}
  \bibinfo{year}{2015}\natexlab{}.
\newblock \showarticletitle{Automatic classification of object code using
  machine learning}.
\newblock \bibinfo{journal}{\emph{Digital Investigation}}  \bibinfo{volume}{14}
  (\bibinfo{year}{2015}), \bibinfo{pages}{S156--S162}.
\newblock


\bibitem[\protect\citeauthoryear{Costin, Zaddach, Francillon, Balzarotti, and
  Antipolis}{Costin et~al\mbox{.}}{2014}]%
        {costin2014large}
\bibfield{author}{\bibinfo{person}{Andrei Costin}, \bibinfo{person}{Jonas
  Zaddach}, \bibinfo{person}{Aur{\'e}lien Francillon}, \bibinfo{person}{Davide
  Balzarotti}, {and} \bibinfo{person}{Sophia Antipolis}.}
  \bibinfo{year}{2014}\natexlab{}.
\newblock \showarticletitle{A Large-Scale Analysis of the Security of Embedded
  Firmwares}. In \bibinfo{booktitle}{\emph{USENIX Security}}.
\newblock


\bibitem[\protect\citeauthoryear{Costin, Zarras, and Francillon}{Costin
  et~al\mbox{.}}{2016}]%
        {costin2016automated}
\bibfield{author}{\bibinfo{person}{Andrei Costin}, \bibinfo{person}{Apostolis
  Zarras}, {and} \bibinfo{person}{Aur{\'e}lien Francillon}.}
  \bibinfo{year}{2016}\natexlab{}.
\newblock \showarticletitle{Automated dynamic firmware analysis at scale: a
  case study on embedded web interfaces}. In
  \bibinfo{booktitle}{\emph{Proceedings of the 11th ACM on Asia Conference on
  Computer and Communications Security}}. ACM.
\newblock


\bibitem[\protect\citeauthoryear{Costin, Zarras, and Francillon}{Costin
  et~al\mbox{.}}{2017}]%
        {costin2017towards}
\bibfield{author}{\bibinfo{person}{Andrei Costin}, \bibinfo{person}{Apostolis
  Zarras}, {and} \bibinfo{person}{Aur{\'e}lien Francillon}.}
  \bibinfo{year}{2017}\natexlab{}.
\newblock \showarticletitle{Towards automated classification of firmware images
  and identification of embedded devices}. In \bibinfo{booktitle}{\emph{IFIP
  International Conference on ICT Systems Security and Privacy Protection}}.
  Springer, \bibinfo{pages}{233--247}.
\newblock


\bibitem[\protect\citeauthoryear{Davidson, Moench, Ristenpart, and
  Jha}{Davidson et~al\mbox{.}}{2013}]%
        {davidson2013fie}
\bibfield{author}{\bibinfo{person}{Drew Davidson}, \bibinfo{person}{Benjamin
  Moench}, \bibinfo{person}{Thomas Ristenpart}, {and} \bibinfo{person}{Somesh
  Jha}.} \bibinfo{year}{2013}\natexlab{}.
\newblock \showarticletitle{{FIE on Firmware: Finding Vulnerabilities in
  Embedded Systems Using Symbolic Execution}}. In
  \bibinfo{booktitle}{\emph{USENIX Security}}.
\newblock


\bibitem[\protect\citeauthoryear{De~Nicolao, Pogliani, Polino, Carminati,
  Quarta, and Zanero}{De~Nicolao et~al\mbox{.}}{2018}]%
        {de2018elisa}
\bibfield{author}{\bibinfo{person}{Pietro De~Nicolao},
  \bibinfo{person}{Marcello Pogliani}, \bibinfo{person}{Mario Polino},
  \bibinfo{person}{Michele Carminati}, \bibinfo{person}{Davide Quarta}, {and}
  \bibinfo{person}{Stefano Zanero}.} \bibinfo{year}{2018}\natexlab{}.
\newblock \showarticletitle{{ELISA: ELiciting ISA of Raw Binaries for
  Fine-Grained Code and Data Separation}}. In
  \bibinfo{booktitle}{\emph{International Conference on Detection of Intrusions
  and Malware, and Vulnerability Assessment}}. Springer,
  \bibinfo{pages}{351--371}.
\newblock


\bibitem[\protect\citeauthoryear{Delugr{\'e}}{Delugr{\'e}}{2010}]%
        {delugre2010closer}
\bibfield{author}{\bibinfo{person}{Guillaume Delugr{\'e}}.}
  \bibinfo{year}{2010}\natexlab{}.
\newblock \showarticletitle{Closer to metal: reverse-engineering the Broadcom
  NetExtreme’s firmware}.
\newblock \bibinfo{journal}{\emph{Hack.Lu}}  \bibinfo{volume}{10}
  (\bibinfo{year}{2010}).
\newblock


\bibitem[\protect\citeauthoryear{Duflot, Perez, and Morin}{Duflot
  et~al\mbox{.}}{2011}]%
        {duflot2011if}
\bibfield{author}{\bibinfo{person}{Lo{\"\i}c Duflot},
  \bibinfo{person}{Yves-Alexis Perez}, {and} \bibinfo{person}{Benjamin Morin}.}
  \bibinfo{year}{2011}\natexlab{}.
\newblock \showarticletitle{What if you can’t trust your network card?}. In
  \bibinfo{booktitle}{\emph{International Workshop on Recent Advances in
  Intrusion Detection}}.
\newblock


\bibitem[\protect\citeauthoryear{Durlak and DuPre}{Durlak and DuPre}{2008}]%
        {durlak2008implementation}
\bibfield{author}{\bibinfo{person}{Joseph~A Durlak} {and}
  \bibinfo{person}{Emily~P DuPre}.} \bibinfo{year}{2008}\natexlab{}.
\newblock \showarticletitle{Implementation matters: A review of research on the
  influence of implementation on program outcomes and the factors affecting
  implementation}.
\newblock \bibinfo{journal}{\emph{American journal of community psychology}}
  \bibinfo{volume}{41}, \bibinfo{number}{3-4} (\bibinfo{year}{2008}),
  \bibinfo{pages}{327--350}.
\newblock


\bibitem[\protect\citeauthoryear{Eagle}{Eagle}{2011}]%
        {eagle2011ida}
\bibfield{author}{\bibinfo{person}{Chris Eagle}.}
  \bibinfo{year}{2011}\natexlab{}.
\newblock \bibinfo{booktitle}{\emph{{The IDA pro book}}}.
\newblock \bibinfo{publisher}{No Starch Press}.
\newblock


\bibitem[\protect\citeauthoryear{Eschweiler, Yakdan, and
  Gerhards-Padilla}{Eschweiler et~al\mbox{.}}{2016}]%
        {eschweiler2016discovre}
\bibfield{author}{\bibinfo{person}{Sebastian Eschweiler},
  \bibinfo{person}{Khaled Yakdan}, {and} \bibinfo{person}{Elmar
  Gerhards-Padilla}.} \bibinfo{year}{2016}\natexlab{}.
\newblock \showarticletitle{{discovRE: Efficient Cross-Architecture
  Identification of Bugs in Binary Code.}}. In
  \bibinfo{booktitle}{\emph{NDSS}}.
\newblock


\bibitem[\protect\citeauthoryear{Feng, Zhou, Xu, Cheng, Testa, and Yin}{Feng
  et~al\mbox{.}}{2016}]%
        {feng2016scalable}
\bibfield{author}{\bibinfo{person}{Qian Feng}, \bibinfo{person}{Rundong Zhou},
  \bibinfo{person}{Chengcheng Xu}, \bibinfo{person}{Yao Cheng},
  \bibinfo{person}{Brian Testa}, {and} \bibinfo{person}{Heng Yin}.}
  \bibinfo{year}{2016}\natexlab{}.
\newblock \showarticletitle{Scalable graph-based bug search for firmware
  images}. In \bibinfo{booktitle}{\emph{Proceedings of the 2016 ACM SIGSAC
  Conference on Computer and Communications Security}}. ACM,
  \bibinfo{pages}{480--491}.
\newblock


\bibitem[\protect\citeauthoryear{Fitzgerald, Mathews, Morris, and
  Zhulyn}{Fitzgerald et~al\mbox{.}}{2012}]%
        {fitzgerald2012using}
\bibfield{author}{\bibinfo{person}{Simran Fitzgerald}, \bibinfo{person}{George
  Mathews}, \bibinfo{person}{Colin Morris}, {and} \bibinfo{person}{Oles
  Zhulyn}.} \bibinfo{year}{2012}\natexlab{}.
\newblock \showarticletitle{Using NLP techniques for file fragment
  classification}.
\newblock \bibinfo{journal}{\emph{Digital Investigation}}  \bibinfo{volume}{9}
  (\bibinfo{year}{2012}), \bibinfo{pages}{S44--S49}.
\newblock


\bibitem[\protect\citeauthoryear{Foster}{Foster}{2005}]%
        {foster2005sockets}
\bibfield{author}{\bibinfo{person}{James~C Foster}.}
  \bibinfo{year}{2005}\natexlab{}.
\newblock \bibinfo{booktitle}{\emph{{Sockets, Shellcode, Porting, and Coding:
  Reverse Engineering Exploits and Tool Coding for Security Professionals}}}.
\newblock \bibinfo{publisher}{Elsevier}.
\newblock


\bibitem[\protect\citeauthoryear{Granboulan}{Granboulan}{[n. d.]}]%
        {cpurec}
\bibfield{author}{\bibinfo{person}{Louis Granboulan}.} \bibinfo{year}{[n.
  d.]}\natexlab{}.
\newblock \bibinfo{title}{{cpu\_rec -- Recognize CPU instructions in an
  arbitrary binary file}}.
\newblock
\newblock
\urldef\tempurl%
\url{https://github.com/airbus-seclab/cpu\_rec}
\showURL{%
\tempurl}


\bibitem[\protect\citeauthoryear{Granboulan}{Granboulan}{2017}]%
        {granboulan2017cpurec}
\bibfield{author}{\bibinfo{person}{Louis Granboulan}.}
  \bibinfo{year}{2017}\natexlab{}.
\newblock \showarticletitle{cpu\_rec.py}.
\newblock \bibinfo{journal}{\emph{SSTIC}} (\bibinfo{year}{2017}).
\newblock
\urldef\tempurl%
\url{https://airbus-seclab.github.io/cpurec/SSTIC2017-Article-cpu_rec-granboulan.pdf}
\showURL{%
\tempurl}


\bibitem[\protect\citeauthoryear{Japkowicz}{Japkowicz}{2003}]%
        {japkowicz2003class}
\bibfield{author}{\bibinfo{person}{Nathalie Japkowicz}.}
  \bibinfo{year}{2003}\natexlab{}.
\newblock \showarticletitle{Class imbalances: are we focusing on the right
  issue}. In \bibinfo{booktitle}{\emph{Workshop on Learning from Imbalanced
  Data Sets II}}, Vol.~\bibinfo{volume}{1723}. \bibinfo{pages}{63}.
\newblock


\bibitem[\protect\citeauthoryear{Kadav and Swift}{Kadav and Swift}{2012}]%
        {kadav2012understanding}
\bibfield{author}{\bibinfo{person}{Asim Kadav} {and} \bibinfo{person}{Michael~M
  Swift}.} \bibinfo{year}{2012}\natexlab{}.
\newblock \showarticletitle{Understanding modern device drivers}.
\newblock \bibinfo{journal}{\emph{ACM SIGARCH Computer Architecture News}}
  \bibinfo{volume}{40}, \bibinfo{number}{1} (\bibinfo{year}{2012}),
  \bibinfo{pages}{87--98}.
\newblock


\bibitem[\protect\citeauthoryear{Kairaj\"arvi}{Kairaj\"arvi}{2019}]%
        {Kairajarvi:Thesis:2019}
\bibfield{author}{\bibinfo{person}{Sami Kairaj\"arvi}.}
  \bibinfo{year}{2019}\natexlab{}.
\newblock \emph{\bibinfo{title}{{Automatic identification of architecture and
  endianness using binary file contents}}}.
\newblock \bibinfo{thesistype}{Master's\ thesis}. \bibinfo{school}{University
  of Jyv\"askyl\"a}, \bibinfo{address}{Jyv\"askyl\"a, Finland}.
\newblock
\urldef\tempurl%
\url{http://urn.fi/URN:NBN:fi:jyu-201904182217}
\showURL{%
\tempurl}


\bibitem[\protect\citeauthoryear{Kriegel, Schubert, and Zimek}{Kriegel
  et~al\mbox{.}}{2017}]%
        {kriegel2017black}
\bibfield{author}{\bibinfo{person}{Hans-Peter Kriegel}, \bibinfo{person}{Erich
  Schubert}, {and} \bibinfo{person}{Arthur Zimek}.}
  \bibinfo{year}{2017}\natexlab{}.
\newblock \showarticletitle{The (black) art of runtime evaluation: Are we
  comparing algorithms or implementations?}
\newblock \bibinfo{journal}{\emph{Knowledge and Information Systems}}
  \bibinfo{volume}{52}, \bibinfo{number}{2} (\bibinfo{year}{2017}),
  \bibinfo{pages}{341--378}.
\newblock


\bibitem[\protect\citeauthoryear{Li, Ong, Suganthan, and Thing}{Li
  et~al\mbox{.}}{2011b}]%
        {li2011novel}
\bibfield{author}{\bibinfo{person}{Qiming Li}, \bibinfo{person}{A Ong},
  \bibinfo{person}{P Suganthan}, {and} \bibinfo{person}{V Thing}.}
  \bibinfo{year}{2011}\natexlab{b}.
\newblock \showarticletitle{A novel support vector machine approach to high
  entropy data fragment classification}. In
  \bibinfo{booktitle}{\emph{Proceedings of the South African Information
  Security Multi-Conf (SAISMC)}}. \bibinfo{pages}{236--247}.
\newblock


\bibitem[\protect\citeauthoryear{Li, Wang, Stolfo, and Herzog}{Li
  et~al\mbox{.}}{2005}]%
        {li2005fileprints}
\bibfield{author}{\bibinfo{person}{Wei-Jen Li}, \bibinfo{person}{Ke Wang},
  \bibinfo{person}{Salvatore~J Stolfo}, {and} \bibinfo{person}{Benjamin
  Herzog}.} \bibinfo{year}{2005}\natexlab{}.
\newblock \showarticletitle{Fileprints: Identifying file types by n-gram
  analysis}. In \bibinfo{booktitle}{\emph{Proceedings from the Sixth Annual
  IEEE SMC Information Assurance Workshop}}. IEEE, \bibinfo{pages}{64--71}.
\newblock


\bibitem[\protect\citeauthoryear{Li, McCune, and Perrig}{Li
  et~al\mbox{.}}{2011a}]%
        {li2011viper}
\bibfield{author}{\bibinfo{person}{Yanlin Li}, \bibinfo{person}{Jonathan~M
  McCune}, {and} \bibinfo{person}{Adrian Perrig}.}
  \bibinfo{year}{2011}\natexlab{a}.
\newblock \showarticletitle{{VIPER: verifying the integrity of PERipherals'
  firmware}}. In \bibinfo{booktitle}{\emph{Proceedings of the 18th ACM
  conference on Computer and communications security}}. ACM,
  \bibinfo{pages}{3--16}.
\newblock


\bibitem[\protect\citeauthoryear{Liu, Tan, and Chen}{Liu et~al\mbox{.}}{2013}]%
        {liu2013binary}
\bibfield{author}{\bibinfo{person}{Kaiping Liu}, \bibinfo{person}{Hee Beng~Kuan
  Tan}, {and} \bibinfo{person}{Xu Chen}.} \bibinfo{year}{2013}\natexlab{}.
\newblock \showarticletitle{Binary code analysis}.
\newblock \bibinfo{journal}{\emph{Computer}} \bibinfo{volume}{46},
  \bibinfo{number}{8} (\bibinfo{year}{2013}).
\newblock


\bibitem[\protect\citeauthoryear{Manni, Aziz, Gong, Loganathan, and Amin}{Manni
  et~al\mbox{.}}{2014}]%
        {manni2014network}
\bibfield{author}{\bibinfo{person}{Jayaraman Manni}, \bibinfo{person}{Ashar
  Aziz}, \bibinfo{person}{Fengmin Gong}, \bibinfo{person}{Upendran Loganathan},
  {and} \bibinfo{person}{Muhammad Amin}.} \bibinfo{year}{2014}\natexlab{}.
\newblock \bibinfo{title}{Network-based binary file extraction and analysis for
  malware detection}.
\newblock
\newblock
\newblock
\shownote{US Patent 8,832,829.}


\bibitem[\protect\citeauthoryear{McDaniel and Heydari}{McDaniel and
  Heydari}{2003}]%
        {mcdaniel2003content}
\bibfield{author}{\bibinfo{person}{Mason McDaniel} {and}
  \bibinfo{person}{Mohammad~Hossain Heydari}.} \bibinfo{year}{2003}\natexlab{}.
\newblock \showarticletitle{Content based file type detection algorithms}. In
  \bibinfo{booktitle}{\emph{36th Annual Hawaii International Conference on
  System Sciences, 2003. Proceedings of the}}. IEEE, \bibinfo{pages}{10--pp}.
\newblock


\bibitem[\protect\citeauthoryear{Miller}{Miller}{2011}]%
        {miller2011battery}
\bibfield{author}{\bibinfo{person}{Charlie Miller}.}
  \bibinfo{year}{2011}\natexlab{}.
\newblock \showarticletitle{Battery firmware hacking}.
\newblock \bibinfo{journal}{\emph{Black Hat USA}} (\bibinfo{year}{2011}),
  \bibinfo{pages}{3--4}.
\newblock


\bibitem[\protect\citeauthoryear{Muench, Stijohann, Kargl, Francillon, and
  Balzarotti}{Muench et~al\mbox{.}}{2018}]%
        {muench2018you}
\bibfield{author}{\bibinfo{person}{Marius Muench}, \bibinfo{person}{Jan
  Stijohann}, \bibinfo{person}{Frank Kargl}, \bibinfo{person}{Aur{\'e}lien
  Francillon}, {and} \bibinfo{person}{Davide Balzarotti}.}
  \bibinfo{year}{2018}\natexlab{}.
\newblock \showarticletitle{What you corrupt is not what you crash: Challenges
  in fuzzing embedded devices}. In \bibinfo{booktitle}{\emph{Proceedings of the
  Network and Distributed System Security Symposium}}.
\newblock


\bibitem[\protect\citeauthoryear{Nohl and Lell}{Nohl and Lell}{2014}]%
        {nohl2014badusb}
\bibfield{author}{\bibinfo{person}{Karsten Nohl} {and} \bibinfo{person}{Jakob
  Lell}.} \bibinfo{year}{2014}\natexlab{}.
\newblock \showarticletitle{{BadUSB-On accessories that turn evil}}.
\newblock \bibinfo{journal}{\emph{Black Hat USA}} (\bibinfo{year}{2014}).
\newblock


\bibitem[\protect\citeauthoryear{"pancake" Alvarez and core
  contributors}{"pancake" Alvarez and core contributors}{[n. d.]}]%
        {radare2}
\bibfield{author}{\bibinfo{person}{Sergi "pancake" Alvarez} {and}
  \bibinfo{person}{core contributors}.} \bibinfo{year}{[n. d.]}\natexlab{}.
\newblock \bibinfo{title}{radare2 -- unix-like reverse engineering framework
  and commandline tools}.
\newblock
\newblock
\urldef\tempurl%
\url{https://www.radare.org/}
\showURL{%
\tempurl}


\bibitem[\protect\citeauthoryear{Pedregosa, Varoquaux, Gramfort, Michel,
  Thirion, Grisel, Blondel, Prettenhofer, Weiss, Dubourg, Vanderplas, Passos,
  Cournapeau, Brucher, Perrot, and Duchesnay}{Pedregosa et~al\mbox{.}}{2011}]%
        {scikit-learn}
\bibfield{author}{\bibinfo{person}{F. Pedregosa}, \bibinfo{person}{G.
  Varoquaux}, \bibinfo{person}{A. Gramfort}, \bibinfo{person}{V. Michel},
  \bibinfo{person}{B. Thirion}, \bibinfo{person}{O. Grisel},
  \bibinfo{person}{M. Blondel}, \bibinfo{person}{P. Prettenhofer},
  \bibinfo{person}{R. Weiss}, \bibinfo{person}{V. Dubourg}, \bibinfo{person}{J.
  Vanderplas}, \bibinfo{person}{A. Passos}, \bibinfo{person}{D. Cournapeau},
  \bibinfo{person}{M. Brucher}, \bibinfo{person}{M. Perrot}, {and}
  \bibinfo{person}{E. Duchesnay}.} \bibinfo{year}{2011}\natexlab{}.
\newblock \showarticletitle{Scikit-learn: Machine Learning in {P}ython}.
\newblock \bibinfo{journal}{\emph{Journal of Machine Learning Research}}
  \bibinfo{volume}{12} (\bibinfo{year}{2011}), \bibinfo{pages}{2825--2830}.
\newblock


\bibitem[\protect\citeauthoryear{Penrose, Macfarlane, and Buchanan}{Penrose
  et~al\mbox{.}}{2013}]%
        {penrose2013approaches}
\bibfield{author}{\bibinfo{person}{Philip Penrose}, \bibinfo{person}{Richard
  Macfarlane}, {and} \bibinfo{person}{William~J Buchanan}.}
  \bibinfo{year}{2013}\natexlab{}.
\newblock \showarticletitle{Approaches to the classification of high entropy
  file fragments}.
\newblock \bibinfo{journal}{\emph{Digital Investigation}} \bibinfo{volume}{10},
  \bibinfo{number}{4} (\bibinfo{year}{2013}), \bibinfo{pages}{372--384}.
\newblock


\bibitem[\protect\citeauthoryear{Pewny, Garmany, Gawlik, Rossow, and
  Holz}{Pewny et~al\mbox{.}}{2015}]%
        {pewny2015cross}
\bibfield{author}{\bibinfo{person}{Jannik Pewny}, \bibinfo{person}{Behrad
  Garmany}, \bibinfo{person}{Robert Gawlik}, \bibinfo{person}{Christian
  Rossow}, {and} \bibinfo{person}{Thorsten Holz}.}
  \bibinfo{year}{2015}\natexlab{}.
\newblock \showarticletitle{Cross-architecture bug search in binary
  executables}. In \bibinfo{booktitle}{\emph{IEEE Symposium on Security and
  Privacy}}.
\newblock


\bibitem[\protect\citeauthoryear{Polychronakis, Anagnostakis, and
  Markatos}{Polychronakis et~al\mbox{.}}{2010}]%
        {polychronakis2010comprehensive}
\bibfield{author}{\bibinfo{person}{Michalis Polychronakis},
  \bibinfo{person}{Kostas~G Anagnostakis}, {and} \bibinfo{person}{Evangelos~P
  Markatos}.} \bibinfo{year}{2010}\natexlab{}.
\newblock \showarticletitle{Comprehensive shellcode detection using runtime
  heuristics}. In \bibinfo{booktitle}{\emph{Proceedings of the 26th Annual
  Computer Security Applications Conference}}. ACM, \bibinfo{pages}{287--296}.
\newblock


\bibitem[\protect\citeauthoryear{Quynh}{Quynh}{2014}]%
        {quynh2014capstone}
\bibfield{author}{\bibinfo{person}{Nguyen~Anh Quynh}.}
  \bibinfo{year}{2014}\natexlab{}.
\newblock \showarticletitle{Capstone: Next-gen disassembly framework}.
\newblock \bibinfo{journal}{\emph{Black Hat USA}} (\bibinfo{year}{2014}).
\newblock


\bibitem[\protect\citeauthoryear{Shoshitaishvili, Wang, Hauser, Kruegel, and
  Vigna}{Shoshitaishvili et~al\mbox{.}}{2015}]%
        {shoshitaishvili2015firmalice}
\bibfield{author}{\bibinfo{person}{Yan Shoshitaishvili}, \bibinfo{person}{Ruoyu
  Wang}, \bibinfo{person}{Christophe Hauser}, \bibinfo{person}{Christopher
  Kruegel}, {and} \bibinfo{person}{Giovanni Vigna}.}
  \bibinfo{year}{2015}\natexlab{}.
\newblock \showarticletitle{{Firmalice-Automatic Detection of Authentication
  Bypass Vulnerabilities in Binary Firmware.}}. In
  \bibinfo{booktitle}{\emph{NDSS}}.
\newblock


\bibitem[\protect\citeauthoryear{Shoshitaishvili, Wang, Salls, Stephens,
  Polino, Dutcher, Grosen, Feng, Hauser, Kruegel,
  et~al\mbox{.}}{Shoshitaishvili et~al\mbox{.}}{2016b}]%
        {shoshitaishvili2016sok}
\bibfield{author}{\bibinfo{person}{Yan Shoshitaishvili}, \bibinfo{person}{Ruoyu
  Wang}, \bibinfo{person}{Christopher Salls}, \bibinfo{person}{Nick Stephens},
  \bibinfo{person}{Mario Polino}, \bibinfo{person}{Andrew Dutcher},
  \bibinfo{person}{John Grosen}, \bibinfo{person}{Siji Feng},
  \bibinfo{person}{Christophe Hauser}, \bibinfo{person}{Christopher Kruegel},
  {et~al\mbox{.}}} \bibinfo{year}{2016}\natexlab{b}.
\newblock \showarticletitle{Sok:(state of) the art of war: Offensive techniques
  in binary analysis}. In \bibinfo{booktitle}{\emph{2016 IEEE Symposium on
  Security and Privacy (SP)}}. IEEE, \bibinfo{pages}{138--157}.
\newblock


\bibitem[\protect\citeauthoryear{Shoshitaishvili, Wang, Salls, Stephens,
  Polino, Dutcher, Grosen, Feng, Hauser, Kruegel, and Vigna}{Shoshitaishvili
  et~al\mbox{.}}{2016a}]%
        {shoshitaishvili2016state}
\bibfield{author}{\bibinfo{person}{Yan Shoshitaishvili}, \bibinfo{person}{Ruoyu
  Wang}, \bibinfo{person}{Christopher Salls}, \bibinfo{person}{Nick Stephens},
  \bibinfo{person}{Mario Polino}, \bibinfo{person}{Audrey Dutcher},
  \bibinfo{person}{John Grosen}, \bibinfo{person}{Siji Feng},
  \bibinfo{person}{Christophe Hauser}, \bibinfo{person}{Christopher Kruegel},
  {and} \bibinfo{person}{Giovanni Vigna}.} \bibinfo{year}{2016}\natexlab{a}.
\newblock \showarticletitle{{SoK: (State of) The Art of War: Offensive
  Techniques in Binary Analysis}}. In \bibinfo{booktitle}{\emph{IEEE Symposium
  on Security and Privacy}}.
\newblock


\bibitem[\protect\citeauthoryear{Sites, Chernoff, Kirk, Marks, and
  Robinson}{Sites et~al\mbox{.}}{1993}]%
        {sites1993binary}
\bibfield{author}{\bibinfo{person}{Richard~L Sites}, \bibinfo{person}{Anton
  Chernoff}, \bibinfo{person}{Matthew~B Kirk}, \bibinfo{person}{Maurice~P
  Marks}, {and} \bibinfo{person}{Scott~G Robinson}.}
  \bibinfo{year}{1993}\natexlab{}.
\newblock \showarticletitle{Binary translation}.
\newblock \bibinfo{journal}{\emph{Digital Technical Journal}}
  \bibinfo{volume}{4} (\bibinfo{year}{1993}), \bibinfo{pages}{137--137}.
\newblock


\bibitem[\protect\citeauthoryear{Song, Brumley, Yin, Caballero, Jager, Kang,
  Liang, Newsome, Poosankam, and Saxena}{Song et~al\mbox{.}}{2008}]%
        {song2008bitblaze}
\bibfield{author}{\bibinfo{person}{Dawn Song}, \bibinfo{person}{David Brumley},
  \bibinfo{person}{Heng Yin}, \bibinfo{person}{Juan Caballero},
  \bibinfo{person}{Ivan Jager}, \bibinfo{person}{Min~Gyung Kang},
  \bibinfo{person}{Zhenkai Liang}, \bibinfo{person}{James Newsome},
  \bibinfo{person}{Pongsin Poosankam}, {and} \bibinfo{person}{Prateek Saxena}.}
  \bibinfo{year}{2008}\natexlab{}.
\newblock \showarticletitle{BitBlaze: A new approach to computer security via
  binary analysis}. In \bibinfo{booktitle}{\emph{International Conference on
  Information Systems Security}}. Springer, \bibinfo{pages}{1--25}.
\newblock


\bibitem[\protect\citeauthoryear{Sportiello and Zanero}{Sportiello and
  Zanero}{2012}]%
        {sportiello2012context}
\bibfield{author}{\bibinfo{person}{Luigi Sportiello} {and}
  \bibinfo{person}{Stefano Zanero}.} \bibinfo{year}{2012}\natexlab{}.
\newblock \showarticletitle{Context-based file block classification}. In
  \bibinfo{booktitle}{\emph{IFIP International Conference on Digital
  Forensics}}. Springer, \bibinfo{pages}{67--82}.
\newblock


\bibitem[\protect\citeauthoryear{Sutherland, Kalb, Blyth, and
  Mulley}{Sutherland et~al\mbox{.}}{2006}]%
        {sutherland2006empirical}
\bibfield{author}{\bibinfo{person}{Iain Sutherland}, \bibinfo{person}{George~E
  Kalb}, \bibinfo{person}{Andrew Blyth}, {and} \bibinfo{person}{Gaius Mulley}.}
  \bibinfo{year}{2006}\natexlab{}.
\newblock \showarticletitle{An empirical examination of the reverse engineering
  process for binary files}.
\newblock \bibinfo{journal}{\emph{Computers \& Security}} \bibinfo{volume}{25},
  \bibinfo{number}{3} (\bibinfo{year}{2006}).
\newblock


\bibitem[\protect\citeauthoryear{Tian, Bates, and Butler}{Tian
  et~al\mbox{.}}{2015}]%
        {tian2015defending}
\bibfield{author}{\bibinfo{person}{Dave~Jing Tian}, \bibinfo{person}{Adam
  Bates}, {and} \bibinfo{person}{Kevin Butler}.}
  \bibinfo{year}{2015}\natexlab{}.
\newblock \showarticletitle{{Defending against malicious USB firmware with
  GoodUSB}}. In \bibinfo{booktitle}{\emph{Proceedings of the 31st Annual
  Computer Security Applications Conference}}. ACM, \bibinfo{pages}{261--270}.
\newblock


\bibitem[\protect\citeauthoryear{Van Den~Berg and Chinchani}{Van Den~Berg and
  Chinchani}{2009}]%
        {van2009detecting}
\bibfield{author}{\bibinfo{person}{Eric Van Den~Berg} {and}
  \bibinfo{person}{Ramkumar Chinchani}.} \bibinfo{year}{2009}\natexlab{}.
\newblock \bibinfo{title}{Detecting exploit code in network flows}.
\newblock
\newblock
\newblock
\shownote{US Patent App. 11/260,914.}


\bibitem[\protect\citeauthoryear{Wang and Shoshitaishvili}{Wang and
  Shoshitaishvili}{2017}]%
        {wang2017angr}
\bibfield{author}{\bibinfo{person}{Fish Wang} {and} \bibinfo{person}{Yan
  Shoshitaishvili}.} \bibinfo{year}{2017}\natexlab{}.
\newblock \showarticletitle{Angr-the next generation of binary analysis}. In
  \bibinfo{booktitle}{\emph{2017 IEEE Cybersecurity Development (SecDev)}}.
  IEEE, \bibinfo{pages}{8--9}.
\newblock


\bibitem[\protect\citeauthoryear{Wang, Wei, Lin, and Zou}{Wang
  et~al\mbox{.}}{2009}]%
        {wang2009intscope}
\bibfield{author}{\bibinfo{person}{Tielei Wang}, \bibinfo{person}{Tao Wei},
  \bibinfo{person}{Zhiqiang Lin}, {and} \bibinfo{person}{Wei Zou}.}
  \bibinfo{year}{2009}\natexlab{}.
\newblock \showarticletitle{{IntScope: Automatically Detecting Integer Overflow
  Vulnerability in X86 Binary Using Symbolic Execution.}}. In
  \bibinfo{booktitle}{\emph{NDSS}}.
\newblock


\bibitem[\protect\citeauthoryear{Xie, Abdullah, and Sulaiman}{Xie
  et~al\mbox{.}}{2013}]%
        {xie2013byte}
\bibfield{author}{\bibinfo{person}{H Xie}, \bibinfo{person}{Azizi Abdullah},
  {and} \bibinfo{person}{Rossilawati Sulaiman}.}
  \bibinfo{year}{2013}\natexlab{}.
\newblock \showarticletitle{Byte frequency analysis descriptor with spatial
  information for file fragment classification}. In
  \bibinfo{booktitle}{\emph{Proceeding of the International Conference on
  Artificial Intelligence in Computer Science and ICT}}.
\newblock


\end{thebibliography}

\section{Appendices}
\label{apdx:apdx}
\appendix

%\section{Additional figures and tables}
%\label{apdx:fig}

\begin{figure}[htb]%{\linewidth}
	\includegraphics[width=\linewidth]{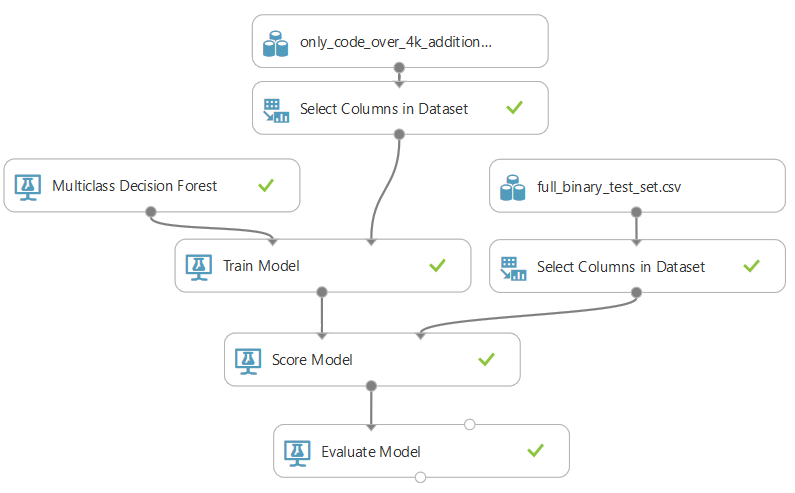}
	\captionof{figure}{An example of setup we used in Azure Machine Learning platform.}
	\label{fig:azure}
\end{figure}

\begin{lstlisting}[label={lst:mlfeatuecalc},caption={Machine learning -- Code to extract and compute feature-sets.},captionpos=t,basicstyle=\tiny,language=Python]
import re

UNKNOWN_ARCHITECTURE = 99
ARCHITECTURES = ['alpha', 'amd64', 'arm64', 'armel', 'armhf', 'hppa', 'i386', 'ia64',
                 'm68k', 'mips', 'mips64el', 'mipsel', 'powerpc', 'powerpcspe',
                 'ppc64', 'ppc64el', 'riscv64', 's390', 's390x', 'sh4', 'sparc',
                 'sparc64', 'x32']

def get_architecture(argument):
    switcher = {
        1: {"architecture": "alpha", "endianness": "little", "wordsize": 64},
        2: {"architecture": "amd64", "endianness": "little", "wordsize": 64},
        [.......]
        23: {"architecture": "x32", "endianness": "little", "wordsize": 32},
        UNKNOWN_ARCHITECTURE: "unknown"
    }
    return switcher.get(argument, "unknown")

class BFD():
    HEADERS_WRITTEN = False

    def initialize_fingerprints(self):
        self.fps = {}
        # big endian one
        self.fps["be_one"] = br"\x00\x01"
        # little endian one
        self.fps["le_one"] = br"\x01\x00"
        # big endian stack
        self.fps["be_stack"] = br"\xff\xfe"
        # little endian stack
        self.fps["le_stack"] = br"\xfe\xff"
        # armel32 prologs
        self.fps["armel32_prolog_1"] = br"[\x00-\xff][\x00-\xff]\x2d\xe9"
        self.fps["armel32_prolog_2"] = br"\x04\xe0\x2d\xe5"
        # armel32 epilogs
        self.fps["armel32_epilog_1"] = br"[\x00-\xff]{2}\xbd\xe8\x1e\xff\x2f\xe1"
        self.fps["armel32_epilog_2"] = br"\x04\xe0\x9d\xe4\x1e\xff\x2f\xe1"
        # arm32 prologs
        self.fps["arm32_prolog_1"] = br"\xe9\x2d[\x00-\xff][\x00-\xff]"
        self.fps["arm32_prolog_2"] = br"\xe5\x2d\xe0\x04"
        # arm32 epilogs
        self.fps["arm32_epilog_1"] = br"\xe8\xbd[\x00-\xff]{2}\xe1\x2f\xff\x1e"
        self.fps["arm32_epilog_2"] = br"\xe4\x9d\xe0\x04\xe1\x2f\xff\x1e"
        # mips32 prologs
        self.fps["mips32_prolog_1"] = br"\x27\xbd\xff[\x00-\xff]"
        self.fps["mips32_prolog_2"] = br"\x3c\x1c[\x00-\xff][\x00-\xff]\x9c\x27" + \
            "[\x00-\xff][\x00-\xff]"
        # mips32 epilog
        self.fps["mips32_epilog_1"] = br"\x8f\xbf[\x00-\xff]{2}([\x00-\xff]{4})" + \
            "{0,4}\x03\xe0\x00\x08"
        # mips32el prologs
        self.fps["mips32el_prolog_1"] = br"[\x00-\xff]\xff\xbd\x27"
        self.fps["mips32el_prolog_2"] = br"[\x00-\xff][\x00-\xff]\x1c\x3c" + \
            "[\x00-\xff][\x00-\xff]\x9c\x27"
        # mipsel epilog
        self.fps["mips32el_epilog_1"] = br"[\x00-\xff]{2}\xbf\x8f([\x00-\xff]" + \
            "{4}){0,4}\x08\x00\xe0\x03"
        # ppc32 prolog
        self.fps["ppc32_prolog_1"] = br"\x94\x21[\x00-\xff]{2}\x7c\x08\x02\xa6"
        # ppc32 epilog
        self.fps["ppc32_epilog_1"] = br"[\x00-\xff]{2}\x03\xa6([\x00-\xff]{4})" + \
            "{0,6}\x4e\x80\x00\x20"
        # ppcel32 prolog
        self.fps["ppcel32_prolog_1"] = br"[\x00-\xff]{2}\x21\x94\xa6\x02\x08\x7c"
        # ppcel32 epilog
        self.fps["ppcel32_epilog_1"] = br"\xa6\x03[\x00-\xff]{2}([\x00-\xff]{4})" + \
            "{0,6}\x20\x00\x80\x4e"
        # ppc64 prologs
        self.fps["ppc64_prolog_1"] = br"\x94\x21[\x00-\xff]{2}\x7c\x08\x02\xa6"
        self.fps["ppc64_prolog_2"] = br"(?!\x94\x21[\x00-\xff]{2})\x7c\x08\x02\xa6"
        self.fps["ppc64_prolog_3"] = br"\xf8\x61[\x00-\xff]{2}"
        # ppc64 epilog
        self.fps["ppc64_epilog_1"] = br"[\x00-\xff]{2}\x03\xa6([\x00-\xff]{4})" + \
            "{0,6}\x4e\x80\x00\x20"
        # ppcel64 prolog
        self.fps["ppcel64_prolog_1"] = br"[\x00-\xff]{2}\x21\x94\xa6\x02\x08\x7c"
        # ppcel64 epilog
        self.fps["ppcel64_epilog_1"] = br"\xa6\x03[\x00-\xff]{2}([\x00-\xff]{4})" + \
            "{0,6}\x20\x00\x80\x4e"
        # s390x prolog
        self.fps["s390x_prolog_1"] = br'\xeb.[\xf0-\xff]..\x24'
        # s390x epilog
        self.fps["s390x_epilog_1"] = br'\x07\xf4'
        # amd64 prologs
        self.fps["amd64_prolog_1"] = br"\x55\x48\x89\xe5"
        self.fps["amd64_prolog_2"] = br"\x48[\x83,\x81]\xec[\x00-\xff]"
        # amd64 epilogs
        self.fps["amd64_epilog_1"] = br"\xc9\xc3"
        self.fps["amd64_epilog_2"] = br"([^\x41][\x50-\x5f]{1}|\x41[\x50-\x5f])\xc3"
        self.fps["amd64_epilog_3"] = br"\x48[\x83,\x81]\xc4([\x00-\xff]{1}|" + \
            "[\x00-\xff]{4})\xc3"
        # powerpcspe SPE instruction
        self.fps["powerpcspe_spe_instruction_isel"] = br"[\x7d-\x7f]" + \
            "[\x00-\xff]{2}(\x1e|\x5e|\x9e)"
        self.fps["powerpcspe_spe_instruction_evl"] = br"(\x10|\x11|" + \
            "\x12|\x13)[\x00-\xff]{2}(\x01 |\xc1 |\xc8 |\xc9 |\xc0 |" + \
            "\xd0 |\xd1 |\xda)"

    def init(self):
        self.initialize_fingerprints()
        # Build regexs out of fingerprints
        for key, value in self.fps.items():
            regex = re.compile(value)
            self.fps[key] = regex
        self.byte_frequencies = []
        self.fingerprints = []

    def calc_features(self, analyze_this):
        byte_count = 0

        # Calculate byte frequency
        byte_frequency_counter = [0] * 256
        for byte in analyze_this:
            byte_count += 1
            byte_frequency_counter[byte] += 1
        try:
            byte_frequency_counter = [(x / byte_count)
                                      for x in byte_frequency_counter]
        except Exception as e:
            print(e)
            return

        # Find matches for function epilog and prolog fingerprints
        fingerprints = {}
        for key, value in self.fps.items():
            i = 0
            for match in value.finditer(analyze_this):
                i += 1
            fingerprints[key] = i / byte_count
        self.byte_frequencies.append(byte_frequency_counter)
        self.fingerprints.append(fingerprints)

    def compose_data(self):
        data = []
        data = data + self.byte_frequencies[0]
        for key in sorted(self.fingerprints[0]):
            data.append(self.fingerprints[0][key])
        return data

def calculate_features(input_file):
    bfd = BFD()
    bfd.init()
    bfd.calc_features(input_file)
    data = bfd.compose_data()
    return data
\end{lstlisting}

%\subsection{Plugins for Reverse Engineering tools}
%\label{apdx:codewsflow}

\begin{lstlisting}[label={lst:radareplugin},caption={Plugin for 
Radare2 for binary under analysis: call our or yours (localhost, public/private IP) web-service API; get the detected architecture, endinanness, wordsize; set the Radare2 internals.},captionpos=t,basicstyle=\tiny,language=Python]

import r2lang, r2pipe, requests

R2P = r2pipe.open()

api_url = "http://localhost:5000/binary/"
api_key = "testkey"

def r2binare(_):
    def process(command):
        if not command.startswith("binare"):
            return 0

        file_size_cmd = "iZ"
        file_size = R2P.cmd(file_size_cmd)

        binary_in_hex_cmd = "p8 %s" % file_size
        binary_in_hex = R2P.cmd(binary_in_hex_cmd)

        data = {"binary": bytearray.fromhex(binary_in_hex.rstrip())}
        form = {"api_key": api_key}
        try:
            response = requests.request(
                "POST", verify=False, url=api_url, files=data, data=form)
        except Exception as e:
            print("Error identifying the architecture")

        response_json = response.json()
        try:
            print("Architecture:", response_json["prediction"]["architecture"],
                  "\nEndianness:", response_json["prediction"]["endianness"],
                  "\nWord size:", response_json["prediction"]["wordsize"],
                  "\nPrediction probabilty:", response_json["prediction_probability"]
                  )
        except Exception:
            print("Unable to identify architecture")

        R2P.cmd("e asm.arch=%s" % response_json["prediction"]["architecture"])
        if response_json["prediction"]["endianness"] == "big":
            R2P.cmd("e cfg.bigendian=true")
        else:
            R2P.cmd("e cfg.bigendian=false")
        R2P.cmd("e asm.bits=%s" % response_json["prediction"]["wordsize"])
\end{lstlisting}

%%%%%%%%%%%%%%%%%%%%%%%%%%%%%%%%%%%%%%%%%%%%%%%%%%%%%%%%%%%%%%%%%%%%%%%%%%%%%%%%

\end{document}